\documentclass[journal]{IEEEtran}
\usepackage{epsfig,graphics,subfigure,psfrag,amsmath,cases}
\usepackage{latexsym,amssymb,amsmath,epsfig,subfigure,algorithm}
\usepackage{algorithmic,graphicx,datetime}
\usepackage{array}
\usepackage{color}
\usepackage{url}
\usepackage{scrtime}

\author{\authorblockN{ Derrick Wing Kwan Ng,~\IEEEmembership{Member,~IEEE,} Ernest S. Lo,~\IEEEmembership{Member,~IEEE,} \\
and Robert Schober,~\IEEEmembership{Fellow,~IEEE}}\\
}

\title{Wireless Information and Power Transfer: Energy Efficiency Optimization in OFDMA Systems
\thanks{Manuscript
received March 16, 2013; revised July 31, 2013; accepted October 1,
2013. The review of this paper was coordinated by Prof. A. Wyglinski.
D. W. K. Ng and R. Schober are with the Institute for Digital Communications (IDC),
Friedrich-Alexander-University Erlangen-N\"urnberg (FAU), Germany (email:\{kwan, schober\}@lnt.de).
Ernest S. Lo is with Centre Tecnol\`{o}gic de Telecomunicacions de
Catalunya - Hong Kong (CTTC-HK) (e-mail: ernest.lo@cttc.hk). This paper has been presented  in part at the IEEE Wireless Communications and Networking Conference (WCNC) 2013 \cite{CN:WCNC_2013} and the IEEE International Conference on Communications (ICC)  2013 \cite{CN:ICC_2013}. This work was supported in part by the AvH Professorship Program of the Alexander von Humboldt Foundation.}
}

\date{\currenttime ,\,\today}

\newtheorem{Thm}{Theorem}

\newcommand{\Rmnum}[1]{\expandafter\@slowromancap\romannumeral #1@}

\newtheorem{Remark}{Remark}

\newcommand{\abs}[1]{\lvert#1\rvert}

\begin{document}
\maketitle
\begin{abstract}
This paper considers orthogonal frequency division multiple access
 (OFDMA) systems with simultaneous wireless information and power transfer.
 We study the resource allocation algorithm design for maximization of
the energy efficiency of data transmission (bits/Joule delivered to
the receivers).  In particular, we focus on power splitting hybrid receivers which are able to split the received signals into two power streams for concurrent information decoding and energy harvesting.  Two scenarios are investigated considering different  power splitting abilities of the receivers. In the first scenario, we assume receivers which can  split the received power into a continuous   set of power streams with  arbitrary power splitting ratios. In the second scenario, we examine receivers which can  split the received power only into a discrete  set of power streams with fixed  power splitting ratios. For both scenarios, we formulate
 the corresponding algorithm design as a
non-convex optimization problem  which takes into account the circuit
power consumption, the minimum data rate requirements of delay constrained services,  the minimum required system data rate, and the minimum amount of power that has to be delivered to the receivers.  By exploiting fractional programming and dual decomposition,
 suboptimal iterative resource allocation algorithms are  developed to solve the  non-convex problems.  Simulation results illustrate that the proposed
iterative resource allocation algorithms approach the optimal solution within a small number of iterations and unveil  the trade-off between energy efficiency, system capacity, and wireless power transfer: (1) wireless power transfer enhances the system energy efficiency by harvesting energy in the radio frequency, especially in the interference limited regime; (2)  the presence of multiple receivers is beneficial for the system capacity, but not necessarily  for the system energy efficiency.

\end{abstract}

\section{Introduction}
\label{sect1}
 \IEEEPARstart{O}{rthogonal} frequency
division multiple access (OFDMA) has been widely adopted as the air interface in high speed wireless
multiuser communication networks, due to its immunity to channel delay spread and flexibility in resource allocation. In an OFDMA system, a wideband  frequency selective spectrum is converted into a number of orthogonal narrowband frequency flat subcarrier
channels  which facilitates the multiplexing of users' data and the exploitation of multiuser diversity.  On the other hand,  next generation communication systems are expected to support multiple users and to guarantee quality of service (QoS). The increasing demand for high data rate and ubiquitous  services  has
led to a high energy consumption in both transmitter(s) and receiver(s). Unfortunately, portable mobile devices are usually only equipped with
limited energy supplies (batteries) which creates bottlenecks in perpetuating the lifetime of networks. Besides, the  battery capacity has improved at a very slow pace over the past decades and is unable to satisfy the new energy requirements \cite{JR:battery}.
Consequently, energy-efficient  mobile communication system design has become a prominent approach for addressing this issue in energy limited networks \cite{CN:WCNC_2013}--\nocite{CN:WCNC_2013,JR:Mag_green,CN:static_power,JR:TCOM_harvesting}\cite{JR:TWC_large_antennas}. Specifically, an enormous number of technologies/methods such as energy harvesting and resource allocation optimization have been proposed in the literature for improving the energy efficiency  (bits-per-Joule) of wireless communication systems. Among the proposed technologies, energy harvesting from the environment is particularly appealing as it constitutes  a perpetual energy source. More importantly,  it provides self-sustainability  to systems  and is virtually free of cost.
\begin{figure*}
 \centering
\includegraphics[width=5in]{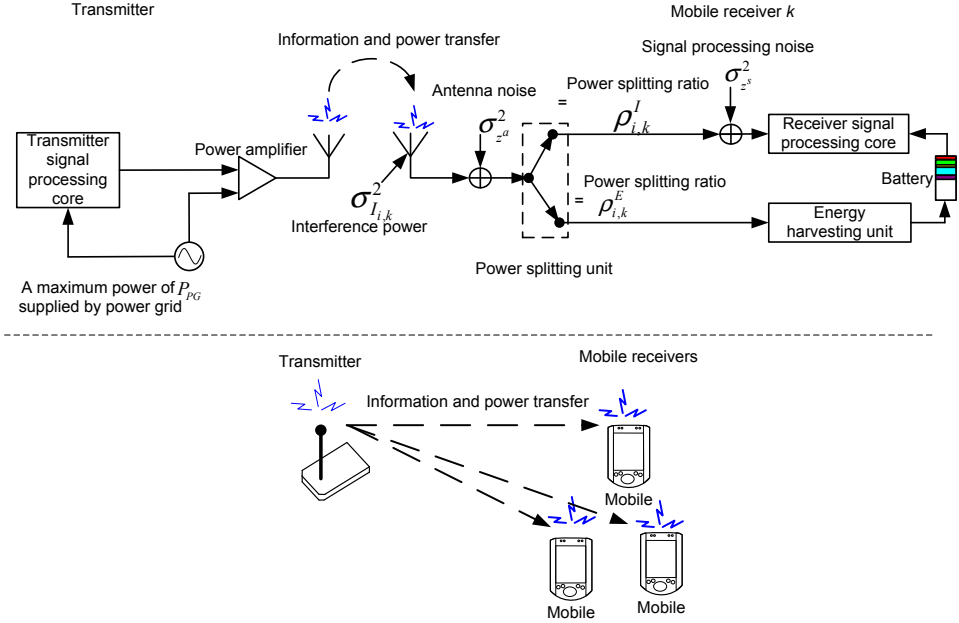}
 \caption{An OFDMA downlink communication system with $K=3$ mobile receivers.  The upper half of the figure illustrates a block diagram of the transceiver model for wireless information and power transfer.} \label{fig:system_model}
\end{figure*}
In practice, numerous renewable energy sources  can
be exploited for energy harvesting, including solar, tide, geothermal,  and wind. However, these  natural energy sources are usually location, weather, or climate dependent and may not always be  available in enclosed/indoor environments or suitable for mobile devices. On the other hand, wireless power transfer technology, which enables the  receivers to scavenge energy  from propagating electromagnetic  waves (EM) in radio frequency (RF),
 has gained recent attention
  in both industry  and academia  \cite{CN:bio_WIP}--\nocite{JR:RFID,CN:WIPT_fundamental,CN:Shannon_meets_tesla,CN:MIMO_WIPT,JR:WIPT_time_switching}\cite{JR:WIP_receiver}. Indeed,   RF signals carry  both information and energy  simultaneously. Thus, the RF energy radiated by the transmitter(s)  can be recycled at the receivers for prolonging the lifetime of networks. Yet, the utilization of EM waves as a carrier for simultaneous information and power transfer poses
  many new research challenges for both  resource allocation algorithm and receiver design. In \cite{CN:WIPT_fundamental} and \cite{CN:Shannon_meets_tesla}, the fundamental trade-off between wireless information and wireless power transfer  was studied for flat fading and frequency selective fading, respectively. Specifically, an ideal receiver was assumed in  \cite{CN:WIPT_fundamental} and \cite{CN:Shannon_meets_tesla} such that information decoding and energy harvesting can be performed on the same received signal which is not possible in practice yet. As a compromise solution, three different types of  receivers, namely \emph{power splitting}, \emph{separated}, and \emph{time-switching} receivers,  were proposed in \cite{CN:MIMO_WIPT,JR:WIPT_time_switching}. In particular, the \emph{power splitting} receiver splits the received power into two power streams with a certain power splitting ratio for facilitating simultaneous  energy harvesting and information decoding in the receiver. The authors in \cite{CN:MIMO_WIPT} and \cite{JR:WIPT_time_switching} investigated the rate-energy regions for different types of receivers  in two-user and point-to-point single carrier systems, respectively.  However, the problem formulations in \cite{CN:MIMO_WIPT} and \cite{JR:WIPT_time_switching} do not take into account the heterogeneous data rate requirements of users  and the solutions may not meet any data rate requirements  and are not applicable to multi-carrier systems with arbitrary numbers of users. On the other hand, the authors in  \cite{JR:WIP_receiver} focused on the resource allocation algorithm design for a point-to-point single user system with \emph{power splitting} receiver in  ergodic fading
channels.  Yet, the assumption of channel ergodicity may not be well justified for delay sensitive services
in practice since the transmitted data symbols of these services
experience slow fading. Besides, the high power consumption in electronic circuitries and  RF transmission  have been overlooked in  \cite{CN:WIPT_fundamental}--\nocite{CN:Shannon_meets_tesla,CN:MIMO_WIPT}\cite{JR:WIP_receiver} but play an important role in designing energy efficient communication systems \cite{JR:Drain_efficiency,JR:energy_efficient_2}.  In other words, the energy efficiency of systems with energy harvesting receivers remains unknown.  In fact, if  a portion of the transmitted RF energy  can be harvested by the RF energy harvesting receivers,   the system energy efficiency may improve. Yet, the system models adopted in \cite{CN:WIPT_fundamental}--\nocite{CN:Shannon_meets_tesla,CN:MIMO_WIPT}\cite{JR:WIP_receiver}  do not consider the energy recycling process from an energy efficiency point of view.  Thus, it is unclear under what conditions energy harvesting receivers improve the system energy efficiency compared to traditional receivers which cannot harvest energy.   By incorporating  the circuit power consumption and the RF energy harvesting ability of the receivers  into the problem formulation, we  presented in  \cite{CN:WCNC_2013} and  \cite{CN:ICC_2013} two energy efficient resource allocation algorithms for multicarrier systems employing \emph{separated} receivers and  \emph{power splitting} receivers, respectively. Nevertheless,  \cite{CN:WCNC_2013} and  \cite{CN:ICC_2013} do not fully exploit the degrees of freedom in resource allocation since  data multiplexing of different users on different subcarriers  was not considered. Moreover, the algorithm proposed in \cite{CN:ICC_2013} incurs a high computation complexity at the transmitter since the optimal power splitting ratio is found via full search over a continuous variable.
Furthermore, the  power splitting ratio may take only discrete levels in practice and the results in  \cite{CN:ICC_2013}, which were obtained for continuous  power splitting ratios, are not applicable in this case.

In this paper, we address the above issues. To this end, we
formulate the resource allocation algorithm design for energy efficient
communication in OFDMA systems with simultaneous wireless information and power transfer
 as an optimization problem. In particular, we focus on the algorithm design for \emph{power splitting} receivers and consider both continuous  and discrete  power splitting ratios. Besides, data  multiplexing of users on different subcarriers is incorporated into the problem formulation.  The resulting non-convex optimization problems are solved by iterative algorithms  which combine nonlinear fractional programming and  dual decomposition. Simulation results illustrate an interesting trade-off between energy efficiency,  wireless power transfer, and multiuser diversity.
%

\section{System Model}
\label{sect2}
In this section,
we present the adopted OFDMA signal model and the model for the hybrid information and  energy harvesting receiver.


\subsection{OFDMA Channel Model}
We consider an OFDMA downlink system in which a transmitter services $K$ mobile receivers. In particular, each mobile receiver is able to decode information and harvest energy from the received radio signals. All transceivers are equipped with a
single antenna, cf.  Figure \ref{fig:system_model}. The total
 system  bandwidth is $\cal B$ Hertz and there are $n_F$
subcarriers.  We focus on quasi-static block fading channels and assume that the
downlink channel gains can be accurately obtained by feedback from the receivers.  The downlink
received symbol at receiver $k\in\{1,\,\ldots,\,K\}$
on subcarrier $i\in\{1,\,\ldots,\,n_F\}$ is
given by
\begin{eqnarray}
Y_{i,k}=\sqrt{P_{i,k}l_k g_k}H_{i,k}X_{i,k}+I_{i,k}+Z_{i,k}^a,
\end{eqnarray}where $X_{i,k}$, $P_{i,k}$, and ${H}_{i,k}$  are the transmitted
data symbol,  the transmitted power, and the  multipath fading coefficient  from
the transmitter to receiver $k$ on subcarrier $i$, respectively.
 $l_k$ and $g_k$ represent the
path loss and shadowing attenuation from the transmitter to receiver $k$, respectively.
 $Z_{i,k}^a$ is the  additive white Gaussian noise (AWGN) originating from the antenna on subcarrier $i$ of receiver $k$.  It is modeled as Gaussian random variable  with zero mean and variance  $\sigma_{z^a}^2$,  cf. Figure \ref{fig:system_model}.
  $I_{i,k}$ is the received aggregate co-channel interference
 on subcarrier $i$ of receiver $k$ with zero mean and variance $\sigma_{I_{i,k}}^2$. $I_{i,k}$  is emitted by unintended transmitters sharing the same frequency channel.

\subsection{Hybrid Information and Energy Harvesting Receiver}
\label{sect:receiver}
In practice, the model of an energy harvesting receiver depends on its specific implementation. For example,  electromagnetic induction and electromagnetic radiation  are able to transfer wireless power \cite{CN:Shannon_meets_tesla,JR:WIP_receiver}. Nevertheless, the associated hardware circuitries  in the receivers and the corresponding energy harvesting efficiency  can be  significantly different. Besides, the signal used for decoding the modulated information cannot be used for harvesting energy due to hardware limitations \cite{JR:WIP_receiver}. In order to isolate the resource allocation algorithm design from the specific hardware
implementation details,  we do not assume a
particular type of energy harvesting receiver.
In this paper,  we focus on receivers which consist of an energy harvesting unit and a conventional signal processing core unit for concurrent  energy  harvesting and  information decoding, cf. Figure \ref{fig:system_model}.   In particular, we adopt a receiver which splits the received signal into two power streams \cite{JR:WIP_receiver} in the  RF front end with power splitting ratios\footnote{Indeed, $\rho^I_{i,k}$ and $\rho^E_{i,k}$ represent the fractions of the received power of user $k$  on subcarrier $i$ used for information decoding and energy harvesting, respectively. Yet, we follow the convention in the literature \cite{JR:WIP_receiver} and adopt the term ``power splitting ratios" in the paper.  } $\rho^I_{i,k}$ and $\rho^E_{i,k}$.  Subsequently, the two power streams with power splitting ratios $\rho^E_{i,k}$ and $\rho^I_{i,k}$ are used for harvesting the energy and decoding the information contained in the signal, respectively. Indeed, by imposing power splitting ratios\footnote{ In this paper,  a perfect passive power splitting unit is assumed; i.e., the power splitting does not incur any power consumption and does not introduce any power loss or signal processing noise to the system. } of $\rho^I_{i,k}=1,\rho^E_{i,k}=0$ and $\rho^I_{i,k}=0,\rho^E_{i,k}=1$, the hybrid receiver reduces to a tradition information receiver or energy harvesting receiver, respectively. Furthermore, we assume that the harvested energy is used to replenish  a rechargeable battery at the receiver. Besides, each
receiver has a fixed power consumption of $P_{C_R}$ Watts which is used for maintaining the  routine operations in the receiver and is independent of the amount of  harvested power.  We note that in practice the receivers may be  powered  by more than one energy source and the harvested energy may be used as a supplement for supporting the energy consumption\footnote{In this paper, the unit of  Joule-per-second is used for energy consumption. Thus,  the terms ``power" and ``energy" are interchangeable.  } of the receivers \cite{CN:hybrid_energy_source}.

\section{Resource Allocation  - Continuous Set of Power Splitting Ratios}\label{sect:forumlation}
In this section,  we consider the resource allocation algorithm design for maximizing the system energy efficiency for the case of a continuous set of power splitting ratios.  The derived  solution provides not only a useful guideline for choosing a suitable number of  discrete power splitting ratios in the power splitting unit, but also
serves as a performance benchmark for the case of discrete power splitting ratios. Now, we define the system energy efficiency by first introducing the weighted system capacity and the power dissipation of the system.
\subsection{System Energy Efficiency}
\label{subsect:Instaneous_Mutual_information}
Assuming the availability of  perfect channel state information (CSI) at the
receiver, the channel capacity between the transmitter  and  receiver $k$ on
subcarrier $i$ with subcarrier bandwidth $W={\cal B}/n_F$  is given by\footnote{The received interference signal $I_{i,k}$ on each subcarrier is treated as AWGN in order to simplify the algorithm design \cite{JR:WIPT_time_switching,JR:interference_model}. }
\begin{eqnarray}\label{eqn:cap}
C_{i,k}&=&W\log_2\Big(1+P_{i,k}\Gamma_{i,k}\Big)\,\,\,\,
\mbox{and}\,\,\,\,\notag\\
\Gamma_{i,k}&=&\frac{\rho_{i,k}^I l_k
g_k\abs{H_{i,k}}^2}{\rho_{i,k}^I(\sigma_{z^a}^2+\sigma_{I_{i,k}}^2)+\sigma_{z^s}^2},
\end{eqnarray}
where $\sigma_{z^s}^2$
is the signal processing noise power at the receiver   and $P_{i,k} \Gamma_{i,k}$ is the received signal-to-interference-plus-noise ratio (SINR) on subcarrier $i$ at receiver $k$.

 The \emph{weighted system capacity} is defined as the aggregate
 number of  bits  delivered to $K$ receivers  and is given by
\begin{eqnarray}
 \label{eqn:avg-sys-goodput} && \hspace*{-5mm} U({\cal P},{\cal S},\mbox{\boldmath$\rho$} )=\sum_{i=1}^{n_F} \sum_{k=1}^{K} \alpha_k s_{i,k} C_{i,k}\quad [\mbox{bits/s}],
\end{eqnarray}
where ${\cal P}=\{ P_{i,k} \ge 0, \forall i,k\}$ is the power allocation policy, ${\cal S}=\{ s_{i,k}\in\{ 0,1\}, \forall i,k\}$ is the subcarrier allocation policy,  and $\mbox{\boldmath$\rho$}=\{ \rho^I_{i,k},\rho^E_{i,k} \ge 0, \forall i,k\}$ is the power splitting policy with variables $\rho^I_{i,k}$ and $\rho^E_{i,k}$ introduced in Section \ref{sect:receiver}. $\alpha_k \ge 0, \forall k$, is a non-negative weight which accounts for the priorities of different receivers and is specified by the application layer. In practice,  proportional fairness and
max-min fairness can be achieved by varying the values of $\alpha_k$ over time\footnote{Optimizing the value of $\alpha_k$ for achieving different system objectives is beyond the scope of this paper. Interested readers may refer to \cite{CN:weight_opt} for further details. }  \cite{JR:Cross_layer_weighted_sum_cap1}.  On the other hand, for facilitating the resource allocation algorithm design, we incorporate    the total power
consumption of the system  into the optimization objective
function.  In particular, the power consumption of the considered system, $U_{TP}({\cal P},{\cal S},\mbox{\boldmath$\rho$})$, consists of five terms and can be expressed as:
\begin{eqnarray}
 \label{eqn:power_consumption}
U_{TP}({\cal P},{\cal S},\mbox{\boldmath$\rho$})\hspace*{-2mm}&=&\hspace*{-2mm}P_{C_T}+ K P_{C_R}+\sum_{k=1}^{K}
 \sum_{i=1}^{n_F}\varepsilon P_{i,k}s_{i,k} \notag\\
 &&-  \sum_{k=1}^{K}Q_{D_k} - \sum_{k=1}^{K}Q_{I_k} \quad [\mbox{Joule/s}], \\
 \mbox{where}\label{eqn:Power_harvested_d}\,\,\,\,Q_{D_k} &=&\notag\\
  &&\hspace*{-1.9cm}\underbrace{\sum_{i=1}^{n_F} \Big(\sum_{j=1}^{K}P_{i,j}s_{i,j}\Big) l_k g_k \abs{H_{i,k}}^2 \eta_k \rho_{i,k}^E}_{\mbox{ Power harvested  from information signal at receiver $k$}}\\ \mbox{and}\,\,\,\,\label{eqn:Power_harvested_s} Q_{I_k} \hspace*{-3mm} &=&\hspace*{-3mm}\underbrace{\sum_{i=1}^{n_F}(\sigma_{z^a}^2+\sigma_{I_{i,k}}^2 )\rho_{i,k}^E \eta_k}_{\hspace*{-10mm} \underset{\mbox{and antenna noise at receiver $k$ }}{ \mbox{ Power harvested  from interference signal}}}\hspace*{-1cm}.
\end{eqnarray}
The first three terms in (\ref{eqn:power_consumption}), i.e., $P_{C_T}+ K P_{C_R}+\sum_{k=1}^{K}
 \sum_{i=1}^{n_F}\varepsilon P_{i,k}s_{i,k}$, represent the power dissipation required for supporting reliable communication.  $P_{C_T}>0$ is the constant  \emph{signal processing circuit  power consumption} in the transmitters, caused by filters, frequency synthesizer, etc., and is independent of the  power radiated by the transmitter. Variables $K P_{C_R}$ and $\sum_{k=1}^{K}
 \sum_{i=1}^{n_F}\varepsilon P_{i,k}s_{i,k}$ denote the total circuit power consumption in the $K$ receivers and the power dissipation
in the power amplifier of the transmitter, respectively.  To model the power inefficiency
of the power amplifier, we introduce a multiplicative constant, $\varepsilon\ge 1$, for the power radiated by the transmitter in (\ref{eqn:power_consumption})  which takes into account  the joint effect of the drain efficiency and  the power amplifier output backoff \cite{JR:Drain_efficiency}. For example,  if $\varepsilon=10$, then $10$ Watts of power are consumed in the power amplifier for every 1 Watt of power radiated in the RF; the wasted power during the power amplification is dissipated as heat. On the other hand, the last two terms in (\ref{eqn:power_consumption}), i.e., $-  \sum_{k=1}^{K}Q_{D_k} - \sum_{k=1}^{K}Q_{I_k}$, represent  the  harvested energy  at the $K$ receivers. The minus sign in front of $\sum_{k=1}^K Q_{D_k}$ in (\ref{eqn:power_consumption}) indicates that a portion of the power radiated in the RF from the transmitter can be harvested by the $K$ receivers. Besides, $0\le\eta_k\le1$ in (\ref{eqn:Power_harvested_d}) and (\ref{eqn:Power_harvested_s}) is a constant which denotes the energy harvesting efficiency of mobile receiver $k$ in converting the received radio signal to electrical energy for storage. In fact, the term $\eta_k  l_k g_k \abs{H_{i,k}}^2\rho^{E}_{i,k}$ in (\ref{eqn:Power_harvested_d}) can be interpreted as a \emph{frequency selective power transfer efficiency} for transferring power from the transmitter to receiver $k$ on subcarrier $i$. Similarly,  the minus sign in front of $\sum_{k=1}^K Q_{I_k}$ in (\ref{eqn:power_consumption}) accounts for the ability of the receivers to harvest energy from interference signals. The weighted \emph{energy efficiency} of the considered system is defined as the
total average number of bits successfully conveyed to the $K$ receivers per Joule consumed energy and is given by
\begin{eqnarray}
 \label{eqn:avg-sys-eff} \hspace*{-8mm}U_{eff}({\cal P},{\cal S},\mbox{\boldmath$\rho$})&=&
 \frac{U_{}({\cal P},{\cal S},\mbox{\boldmath$\rho$})}{U_{TP}({\cal P},{\cal S},\mbox{\boldmath$\rho$})}\quad [\mbox{bits/Joule}].
\end{eqnarray}

It is interesting to note that the weighted energy efficiency is a quasi-concave function with respect to the power allocation variables, cf.  Appendix A. The quasi-concavity of $U_{eff}({\cal P},{\cal S},\boldsymbol \rho)$ can be exploited to show the quasi-concavity of a transformed version of the resource allocation problem, cf. Section III-B, Remark 2. Furthermore, strong interference  can act as an energy source which supplies energy to the receivers and facilitates  energy savings in the system. This effect is not captured by other system models used in the literature \cite{JR:interference_model,JR:multicell}. Therefore, unlike systems with non-energy harvesting receivers, systems with energy harvesting receivers may benefit from interference  as far as energy efficiency is concerned. The energy efficiency gain due to the introduction of energy harvesting receivers will be evaluated  in the simulation section. On the other hand, we emphasize  that the objective function of the considered problem formulation does not capture how the receivers utilize the harvested energy. For instance, if the receivers use the harvested energy for extending their lifetime, an additional system capacity gain may be achieved since the receivers have more time to receive information. In other cases, the receivers may use the harvested energy for uplink transmission which also results in a system capacity gain.

\begin{Remark}\emph{
Mathematically, it is possible that  $U_{TP}({\cal P},{\cal S}, \boldsymbol\rho)$  takes a negative value. Yet, $U_{TP}({\cal P},{\cal S}, \mbox{\boldmath$\rho$})\hspace*{-1mm}>0$  always holds for the considered system due to the following reasons. First, it can be observed that $\sum_{i=1}^{n_F}\sum_{k=1}^{K}\varepsilon P_{i,k}s_{i,k}\ge \sum_{i=1}^{n_F} \sum_{k=1}^{K} P_{i,k}s_{i,k}> \sum_{k=1}^{K}Q_{D_k}$, where the strict inequality is due to the second law of thermodynamics  from physics.  In particular, the communication channel between the transmitter and the $K$ receivers is a passive system which does not introduce extra energy. Besides,  the energy of the desired  signal  received at the receiver  is  attenuated  due to path loss  and energy scavenging inefficiency. Second,  we consider a system where the transmitter and the receivers consume non-negligible amounts of power \cite{JR:energy_efficient_2} for signal processing such that $P_{C_T} + K P_{C_R} >\sum_{k=1}^{K}Q_{I_k}$.}
\end{Remark}

\subsection{Optimization Problem Formulation}
\label{sect:cross-Layer_formulation}
The optimal power allocation policy, ${\cal P}^*$,  subcarrier allocation policy, ${\cal S}^*$, and power splitting policy  ${\boldsymbol \rho}^*$,  can be
obtained by solving the following optimization problem:
\begin{eqnarray}
\label{eqn:cross-layer}&&\hspace*{5mm} \max_{{\cal P},\,{\cal S},\, {\boldsymbol \rho}}\,\,\, U_{eff}({\cal P},{\cal S},\mbox{\boldmath$\rho$})\\
\notag \mbox{s.t.}&&\mbox{C1: }
Q_{D_k}+Q_{I_k}\ge P_{\min_k}^{\rm req},\forall k,\\
&&\mbox{C2:}\notag\sum_{i=1}^{n_F}\sum_{k=1}^{K}P_{i,k}s_{i,k}\le P_{\max},  \\
&&\hspace*{-0mm}\notag \mbox{C3:}\,\,P_{C_T}+\sum_{i=1}^{n_F}\sum_{k=1}^{K}\varepsilon P_{i,k}s_{i,k}\le P_{PG},\\
&& \notag\mbox{C4:}\sum_{i=1}^{n_F}\sum_{k=1}^{K}s_{i,k} C_{i,k}\ge R_{\min},\\
&&\notag\hspace*{-0mm}\mbox{C5:}\,\, \sum_{i=1}^{n_F}s_{i,k} C_{i,k}\ge R_{\min_k}, \forall k\in {\cal D},\\
&&\notag\notag\mbox{C6:}\,\, P_{i,k}\ge 0, \,\, \forall i,k,\hspace*{5.7mm} \mbox{C7:}\,\, s_{i,k}\in\{0,1\},\forall i,k,\notag\end{eqnarray}\begin{eqnarray}
&&\notag\mbox{C8:}\,\, \sum_{k=1}^{K} s_{i,k}\le 1, \,\, \forall i,\quad \mbox{C9:}\,\, \rho^E_{L}\le\rho^E_{i,k}\le \rho^E_{U},\forall i,k,\\
&&\notag\hspace*{-0.18cm}\mbox{C10:}\,\, \rho^I_{L}\le\rho^I_{i,k}s_{i,k}\le \rho^I_{U} ,\forall i,k,\\
&&\hspace*{-10.2mm} \mbox{C11:}\,\, \rho^I_{i,k} s_{i,k}+ \rho^E_{i,k}\le 1, \,\, \forall i,k,\hspace*{2.5mm}
\mbox{C12:}\,\, \rho^E_{i,k} = \rho^E_{j,k}, \,\, \forall k, i\ne j.\notag
\end{eqnarray}
Variable $P_{\min_k}^{\rm req}$ in C1 is a constant which   specifies the minimum required
 power transfer to  receiver $k$. The value of $P_{\max}$ in C2 puts an upper limit on the power radiated  by the transmitter. The value of $P_{\max}$ is a constant which depends on  the hardware limitations  of the power amplifier. C3 limits the maximum  power supplied by the power grid for supporting the power consumption of the transmitter to $P_{PG}$, cf. Figure \ref{fig:system_model}.
$R_{\min}$  in C4  is the
minimum required data rate of the system. Although $R_{\min}$ is not an optimization variable in this paper,  we can strike a balance between the system energy efficiency and the total system throughput  by varying its value. In particular, when $R_{\min}$ is increasing, the resource allocator may increase the transmit power for satisfying the higher data rate requirement by sacrificing the system energy efficiency. C5 is the minimum required data rate $R_{\min_k}$ for
the \emph{delay constrained} services of receiver $k$,  and is specified by the application layer, and $\cal D$ denotes a set of receivers having \emph{delay constrained} services. C6 is the non-negative orthant constraint for the power allocation variables.
 C7 and C8 indicate  that each subcarrier can be allocated to at most one receiver exclusively in conveying information; inter-user interference is avoided in the system. Besides, C8 indicates that some subcarriers can be excluded from the subcarrier selection process  for energy efficiency maximization. Boundary variables $\rho_{U}^{E}$ and  $\rho_{L}^{E}$ in C9 denote the constant upper  and lower bounds of the power splitting ratio for  harvesting energy, respectively.  The bounds are used to account for the limited capability of the  receivers in splitting the received power. Similarly,  $\rho_{U}^{I}$ and  $\rho_{L}^{I}$ in C10 denote the constant upper and lower bounds of the power splitting ratio for  information decoding where  $\rho_{U}^{E}+\rho_{L}^{I}=1$ and $\rho_{L}^{E}+\rho_{U}^{I}=1$.    C11 is the power splitting constraint of the hybrid information and energy harvesting receivers. Its physical meaning is that the power splitting unit, see Figure \ref{fig:system_model}, is a passive device and no extra power gain  can be achieved by  the power splitting process. C12  constrains  the power spitting ratio  for energy harvesting, $\rho^E_{i,k}$, such that it is identical for all subcarriers in each receiver.  Theoretically, $\rho^E_{i,k}$ can be different across different subcarriers. However, in this case, an analog adaptive passive frequency selective power splitter is required  which results in a high system complexity. Therefore, we consider the more practical scenario where $\rho^E_{i,k} = \rho^E_{j,k}, \forall k,\forall j\ne i$.  We do not explicitly impose $\rho^I_{i,k} = \rho^I_{j,k},\forall k,\forall j\ne i$, in the problem formulation. We note that it can be deduced that if both subcarrier $i$ and subcarrier $j$ are allocated to user $k$ for information decoding, $\rho^I_{i,k}=\rho_{j,k}^I$ will hold eventually due to the constraint $\rho^E_{i,k}=\rho_{j,k}^E,\forall k,j\ne i$. However, imposing $\rho^I_{i,k}=\rho_{j,k}^I,\forall k,j\ne i$ explicitly at the very beginning in the problem formulation makes the development for an efficient solution of the optimization problem more cumbersome.


\subsection{Solution of the Optimization Problem } \label{sect:solution}
The key challenge in solving (\ref{eqn:cross-layer}) is the lack of convexity of the problem formulation. In particular, although the objective function in (\ref{eqn:cross-layer}) is quasi-concave with respect to the power allocation variables,
the objective function is a jointly  non-convex function with respect to $\{{\cal P},{\cal S},\boldsymbol \rho\}$. Besides, constraints C1--C12 do not span a convex solution set due to the integer constraint for subcarrier allocation in C7 and the coupled optimization variables in C1. In general, there is no standard approach for solving non-convex optimization problems. In the extreme case,  an exhaustive search or branch-and-bound method is needed to obtain the global optimal solution which is computationally infeasible even for small $K$ and $n_F$.  In order to make the problem tractable, we transform the objective function and approximate the transformed objective function  in order to simplify the problem. Subsequently, we use the constraint relaxation approach for handling the integer constraint C7 to obtain a close-to-optimal resource allocation algorithm. Next, we
introduce the  objective function transformation via a parametric approach from nonlinear fractional programming.

\subsection{Transformation of the Objective Function} \label{sect:solution_dual_decomposition}

  For notational simplicity, we define $\mathcal{F}$ as the set of
feasible solutions\footnote{We assume that the set is non-empty and compact.} of the optimization problem in
(\ref{eqn:cross-layer}) spanned by constraints C1--C12.  Without loss of generality,  we assume that $\{{\cal P},{\cal S},{\boldsymbol \rho}\}\in\mathcal{F}$ and
  we denote $q^*$ as the
maximum energy efficiency of the considered system which can be expressed as
\begin{eqnarray}
q^*=\frac{U({\cal P^*},{\cal S^*},{\boldsymbol \rho^*})}{U_{TP}({\cal
P^*},{\cal
S^*},{\boldsymbol \rho^*})}=\max_{{\cal P},{\cal S},{\boldsymbol \rho}}\,\frac{U({\cal P},{\cal S},\mbox{\boldmath$\rho$})}{U_{TP}({\cal P},{\cal S}, {\boldsymbol \rho})}.
\end{eqnarray}

Now, we introduce the following important Theorem which is borrowed from nonlinear fractional programming  \cite{JR:fractional} for solving the  optimization problem in (\ref{eqn:cross-layer}).
\begin{Thm}\label{Thm:1}
 The resource allocation policy achieves the maximum energy efficiency $q^*$ if and only if
\begin{eqnarray}\notag
\max_{{\cal P},{\cal S}, {\boldsymbol \rho}}&& \hspace*{-2mm}\,U({\cal
P},{\cal S}, {\boldsymbol \rho})-q^*U_{TP}({\cal P},{\cal S},{\boldsymbol \rho})\\
 =&& \hspace*{-2mm}U({\cal
P^*},{\cal S^*},{\mbox{\boldmath$\rho$}}^*)-q^*U_{TP}({\cal P^*},{\cal S^*}, {\mbox{\boldmath$\rho$}}^*)=0, \label{eqn:thm1}
\end{eqnarray}
\end{Thm}
for $U({\cal P},{\cal S}, { {\boldsymbol \rho}})\ge0$ and $U_{TP}({\cal P},{\cal S}, { {\boldsymbol \rho}})>0$.

 \emph{\,Proof:} Please refer to \cite[Appendix A]{JR:TWC_large_antennas} for a proof of Theorem 1.

Theorem \ref{Thm:1} provides a necessary and sufficient condition in describing the optimal resource allocation policy. In particular,
for an optimization problem
with an objective function in fractional form, there exists an
equivalent optimization problem with an
objective function in subtractive form, e.g. $U({\cal P},{\cal S}, { {\boldsymbol \rho}})-q^*U_{TP}({\cal P},{\cal S}, { {\boldsymbol \rho}})$ in the considered
case,  such  that both problem formulations lead to  the same optimal resource allocation policy. Moreover, the optimal resource allocation policy  will enforce the equality in (\ref{eqn:thm1})  which provides an indicator for verifying the optimality of the solution.  As a result, we can focus on the equivalent objective function and design a resource allocation policy for satisfying Theorem 1 in the rest of the paper.

\subsection{Iterative Algorithm for Energy Efficiency Maximization}
In this section, an iterative algorithm (known as the
Dinkelbach method\footnote{We note that the Dinkelbach method is an application of Newton's method  for root finding, please refer to \cite{JR:Drain_efficiency,JR:Fractional_programing}  for details.} \cite{JR:fractional}) is proposed for solving
(\ref{eqn:cross-layer}) with an equivalent objective function such that the obtained solution satisfies the conditions stated in Theorem 1. The
proposed algorithm is summarized in Table \ref{table:algorithm} (on the next page) and
the convergence to the optimal energy efficiency is guaranteed if the inner problem (\ref{eqn:inner_loop}) is solved in each iteration.

\emph{Proof: }Please refer to \cite[Appendix B]{JR:TWC_large_antennas}  for a proof of convergence.

As shown in Table \ref{table:algorithm},   we solve the following optimization problem for a given
parameter $q$ in each iteration in the
main loop, i.e., lines 3--12:
\begin{eqnarray}\label{eqn:inner_loop}
&&\max_{{\cal P},{\cal S}, {\boldsymbol \rho}}
\quad\,{U}({\cal P},{\cal S}, {\boldsymbol \rho})-q{U}_{TP}({\cal P},{\cal S}, { {\boldsymbol \rho}}
)\nonumber\\
&&\hspace*{-0mm}\mbox{s.t.} \,\,\mbox{C1 -- C12}.
\end{eqnarray}
\begin{table}[t]\caption{Iterative Resource Allocation Algorithm.}\label{table:algorithm}
\begin{algorithm} [H]                    
\renewcommand\thealgorithm{}
\caption{Iterative Resource Allocation Algorithm [Dinkelbach method] }          
\label{alg1}                           
\begin{algorithmic} [1]
\normalsize           
\STATE Initialization: $L_{max}=$ the maximum number of iterations  and  $\Delta=$ the
maximum tolerance
 \STATE Set $q=0$ and iteration index $j=0$

\REPEAT [Iteration Process: Main Loop]
\STATE For
a  given $q$, obtain an intermediate resource allocation policy $\{{\cal P'},{\cal S'}, \boldsymbol \rho'\}$ by solving the problem in (\ref{eqn:inner_loop})
\IF {$U({\cal P'},{\cal S'}, {\boldsymbol \rho'})-q U_{TP}({\cal P'},{\cal S'}, {\cal\boldsymbol \rho'})<\Delta$} \STATE  $\mbox{Convergence}=\,$\TRUE \RETURN
$\{{\cal P^*,\cal S^*,\mbox{\boldmath$\rho$}^*}\}=\{{\cal P',\cal S', \boldsymbol \rho'}\}$ and $q^*=\frac{U({\cal
P'},{\cal S'},\boldsymbol \rho')}{ U_{TP}({\cal P'},{\cal S'},\boldsymbol \rho')}$
 \ELSE \STATE
Set $q=\frac{U({\cal P'},{\cal S'},  {\cal\boldsymbol \rho'})}{ U_{TP}({\cal
P'},{\cal S'},{\cal\boldsymbol \rho'})}$ and $j=j+1$ \STATE  Convergence $=$ \FALSE
 \ENDIF
 \UNTIL{Convergence $=$ \TRUE $\,$or $j=L_{max}$}

\end{algorithmic}
\end{algorithm}

\end{table}We note that for any value of $q$ generated   by Algorithm I in each iteration,  ${U}({\cal P},{\cal S}, {{\mbox{\boldmath$\rho$}}})-q{U}_{TP}({\cal P},{\cal S}, { {\mbox{\boldmath$\rho$}}}
)\ge 0$ is always valid; negative energy efficiencies will not occur.    Please refer  to \cite[Proposition 3]{JR:TWC_large_antennas} for a proof.  In fact, the transformed objective function, i.e.,   ${U}({\cal P},{\cal S}, { {\mbox{\boldmath$\rho$}}})-q{U}_{TP}({\cal P},{\cal S}, { {\mbox{\boldmath$\rho$}}}
)$, has an interesting pricing interpretation from the field of economics.  In particular, ${U}({\cal P},{\cal S}, { {\mbox{\boldmath$\rho$}}})$ indicates the system profit due to information transmission while $q{U}_{TP}({\cal P},{\cal S}, { {\mbox{\boldmath$\rho$}}})$ represents the associated  cost due to energy consumption. Besides, the terms $Q_{D_k}$ and $Q_{I_k}$ in  ${U}_{TP}({\cal P},{\cal S}, { {\mbox{\boldmath$\rho$}}})$  are the corresponding rebate and discount  to the energy cost via energy harvesting, respectively. The optimal value of $q$ represents a scaling factor for balancing  profit and cost.


\subsubsection{Solution of the Main Loop Problem (\ref{eqn:inner_loop})}
The transformed problem has an objective function in subtractive form and is parameterized by variable $q$. Unfortunately, there are still two obstacles in tackling the problem.  First, the power splitting variables for information decoding and energy harvesting, i.e., $\rho^I_{i,k}$  and $\rho^E_{i,k} $, are coupled with the power allocation variables in both the objective function and constraint C1 which complicates the solution.  Second, the combinatorial constraint C7 on the subcarrier allocation variables creates a disjoint feasible solution set which is a hurdle for solving the problem via tools from convex optimization. In order to derive a tractable  resource allocation algorithm,
 we approximate the transformed objective function in the following. First, we approximate the \emph{weighted system capacity} as
\begin{eqnarray}\label{eqn:cap-approx}
 U({\cal P},{\cal S},\mbox{\boldmath$\rho$} )&=&\sum_{i=1}^{n_F} \sum_{k=1}^{K} \alpha_k s_{i,k}W\log_2\Big(1+P_{i,k}\Gamma_{i,k}\Big)\notag
\\
&\approx& \widehat {U}({\cal P},{\cal S},\mbox{\boldmath$\rho$} )\\
 \mbox{where    }  \widehat {U}({\cal P},{\cal S},\mbox{\boldmath$\rho$} )&=&\sum_{i=1}^{n_F} \sum_{k=1}^{K} \alpha_k s_{i,k}W\log_2\Big(P_{i,k}\Gamma_{i,k}\Big)
\end{eqnarray}
which is a tight approximation  for high SINR\footnote{The high SINR assumption is used to bring the optimization problem into convex form which makes the problem mathematically tractable. Indeed,  in the proposed problem formulation, a minimum required system data rate $R_{\min}$ is set to guarantee a desired system  data rate. Furthermore, we focus on a indoor line of sight communication environment, see Section V.  Thus, it is unlikely that a selected receiver has  low SINR.}, i.e., $P_{i,k}\Gamma_{i,k}\gg 1$. On the other hand, we adopt a lower bound on $U_{TP}({\cal P},{\cal S},\mbox{\boldmath$\rho$})$ in the transformed objective function in (\ref{eqn:inner_loop}):
\begin{eqnarray}\notag\label{eqn:U_TP-bound}
&&U_{TP}({\cal P},{\cal S},\mbox{\boldmath$\rho$})\\
&\ge& \widehat {U}_{TP}({\cal P},{\cal S},\mbox{\boldmath$\rho$})=P_{C_T}+ K P_{C_R}+
 \sum_{i=1}^{n_F} \sum_{k=1}^{K}\varepsilon P_{i,k}s_{i,k} \notag\\
 &-& \sum_{k=1}^{K}\sum_{i=1}^{n_F} \Big(\sum_{j=1}^{K}P_{i,j}s_{i,j}\Big) l_k g_k \abs{H_{i,k}}^2\eta_k - \sum_{k=1}^{K}Q_{I_k}.
 \end{eqnarray}
 where $\widehat {U}_{TP}({\cal P},{\cal S},\mbox{\boldmath$\rho$})$ is obtained by setting $\rho_{i,k}^E =1$ in  $U_{TP}({\cal P},{\cal S},\mbox{\boldmath$\rho$})$. Indeed, $\widehat {U}_{TP}({\cal P},{\cal S},\mbox{\boldmath$\rho$})$ can be interpreted as  the use of a  theoretical receiver which is able to fully recycle and harvest the energy of the signal used for  information  decoding\footnote{Note that $\rho_{i,k}^E =1$ is only applied to the received energy harvested from the information signal, but not to the portion harvested from the interference signals.}. As a result, the transformed objective function can be approximated by
\begin{eqnarray}\label{eqn:objective_functions}
&&{U}({\cal P},{\cal S}, { {\mbox{\boldmath$\rho$}}})-q{U}_{TP}({\cal P},{\cal S}, { {\mbox{\boldmath$\rho$}}}
)\notag\\
\lesssim&&  \widehat{U}({\cal P},{\cal S},\mbox{\boldmath$\rho$} )-q \widehat {U}_{TP}({\cal P},{\cal S},\mbox{\boldmath$\rho$}).
\end{eqnarray}
 On the other hand, by exploiting constraint C12,  we can rewrite constraint C1 in the following form to remove the associated  non-convexity:
\begin{eqnarray}\label{eqn:equivalent_contraint_c1}
\mbox{C1: }\hspace*{-1mm}&&\sum_{i=1}^{n_F} \Big(\sum_{j=1}^{K}{P}_{i,j}s_{i,j}\Big) l_k g_k \abs{H_{i,k}}^2 \eta_k\notag\\
 &&+ \sum_{i=1}^{n_F}\sigma_{I_{i,k}}^2  \eta_k \ge \frac{P_{\min_k}^{\rm req}}{\rho_{1,k}^E}, \forall k,
\end{eqnarray}
where the right hand side of the inequality in (\ref{eqn:equivalent_contraint_c1}) is due to constraint $\mbox{C12: } \rho_{1,k}^E=\rho_{2,k}^E=\ldots=\rho_{i,k}^E=\ldots=\rho_{n_F,k}^E$. Besides,   we can further  simplify the algorithm design by replacing C12 with the following equivalent constraint:
 \begin{eqnarray}
  &&\mbox{C13:}\,\, \rho^E_{1,k} = \rho^E_{r,k}, \,\, \forall k, r\in\{2,\ldots,n_F\}.
 \end{eqnarray}

 Next, we handle the combinatorial constraint in C7 by time-sharing relaxation \cite{JR:Roger_OFDMA,JR:Time_sharing_wei_yu}.
 In particular, we relax the subcarrier selection variable $ s_{i,k}$ to be a real value between zero and one instead of a Boolean, i.e., C7: $0\le  s_{i,k} \le 1$. As a result,  $s_{i,k}$ can be interpreted as a time-sharing factor in allocating subcarrier $i$ to $K$ receivers for delivering information. In addition,  we introduce two new auxiliary  variables and define them as $\tilde{P}_{i,k}=P_{i,k}s_{i,k}$ and
 $\tilde{\rho}^I_{i,k}={\rho}^I_{i,k}s_{i,k}$. They represent the actual transmitted power and the power splitting ratio for information decoding on subcarrier $i$ for receiver $k$  under the time-sharing condition, respectively.    On the other hand, we replace $C_{i,k}$ in C4 and C5 in (\ref{eqn:inner_loop}) by $\tilde{C}_{i,k}=W\log_2\Big(\frac{\tilde{P}_{i,k}}{s_{i,k}}\tilde{\Gamma}_{i,k}\Big)$, $\tilde{\Gamma}_{i,k}=\Gamma_{i,k}|_{{\rho}^I_{i,k}=\frac{\tilde{\rho}^I_{i,k}}{s_{i,k}}}$ while  $C_{i,k}\ge \tilde{C}_{i,k}$.
 We note that although $C_{i,k} \ge \tilde{C}_{i,k}$ generally holds,   $C_{i,k} \approx \tilde{C}_{i,k}$ is asymptotically tight in the high SINR regime.  Since the feasible solution set of the problem with $ \tilde{C}_{i,k}$  in C4 and C5  is a subset of the original problem \footnote{In general, the  constraint relaxation used in C7  may result in a superset of the feasible solution set. Yet, it will be shown in the next section that the optimal subcarrier allocation policy with respect to the approximated objective function takes values of either zero or one on each subcarrier. In other words, the subcarrier allocation policy is  Boolean   even though it is allowed to take any real value between zero and one; i.e.,  the size of the feasible solution set does not change with the constraint relaxation in C7.  }, the solution obtained  with the  Dinkelbach method for the approximated objective function can be used as a suboptimal solution to the original optimization problem in (\ref{eqn:cross-layer}). Nevertheless, it will be shown in the simulation section that the proposed resource allocation algorithm achieves a close-to-optimal performance in high SINR.   Note that by considering the transformed optimization problem with the approximated objective function and relaxed constraint C7 in each iteration of the Dinkelbach method, the proposed scheme  converges to the optimal solution of the problem with the  approximated objective function introduced in (\ref{eqn:U_TP-bound}).

\begin{Remark}\emph{
There is an alternative approach for solving the optimization problem in (\ref{eqn:cross-layer}). Instead of using ${U}({\cal P},{\cal S}, { {\mbox{\boldmath$\rho$}}})-q{U}_{TP}({\cal P},{\cal S}, { {\mbox{\boldmath$\rho$}}}
)\lesssim  \widehat{U}({\cal P},{\cal S},\mbox{\boldmath$\rho$} )-q \widehat {U}_{TP}({\cal P},{\cal S},\mbox{\boldmath$\rho$})$ from (\ref{eqn:objective_functions}) in each iteration, we could exploit the fact that
 \begin{eqnarray}\notag
\frac{ {U}({\cal P},{\cal S},\boldsymbol\rho )}{  {U}_{TP}({\cal P},{\cal S},\boldsymbol\rho )}\lesssim
\frac{ \widehat{U}({\cal P},{\cal S},\boldsymbol\rho )}{  \widehat{U}_{TP}({\cal P},{\cal S},\boldsymbol\rho )}.
\end{eqnarray}
 Then, we can use a similar approach as in Appendix A to prove that $\frac{ \widehat{U}({\cal P},{\cal S},\boldsymbol\rho )}{  \widehat{U}_{TP}({\cal P},{\cal S},\boldsymbol\rho )}$ is jointly quasi-concave with respect to $({\cal P},{\cal S},\boldsymbol\rho )$ under the time-sharing relaxation. As a result,  a unique global optimal solution exist and can be obtained by using the bisection method \cite{book:convex}. We note that both approaches achieve the same optimal value. }
\end{Remark}

 The problem with transformed objective function in (\ref{eqn:objective_functions}) is jointly concave with respect to (w.r.t.) all  optimization variables (cf. Appendix B). Besides, it can be further verified that the primal problem satisfies   Slater's constraint qualification. As a result,  strong duality holds and  solving the dual
problem is equivalent for solving the primal problem \cite{book:convex}.  Motivated by this fact, we solve the primal problem by solving its dual problem in the following.
 \subsection{Dual Problem Formulation}
In this subsection,  the resource allocation policy is derived
 via solving the dual problem of (\ref{eqn:inner_loop}) with approximated objective function.
 For this purpose, we first need
the Lagrangian function of the primal problem. The Lagrangian of  (\ref{eqn:inner_loop}) is given by
\begin{eqnarray}\hspace*{-4mm}&&\hspace*{-4mm}{\cal
L}( \mathbf{w}, \lambda,\gamma,\beta,\mbox{\boldmath$\delta$},\boldsymbol{\upsilon}
,\mbox{\boldmath$\mu$},\mbox{\boldmath$\zeta$},{\cal P},{\cal S},{\mbox{\boldmath$\rho$}})\\
\notag&=&\hspace*{-3mm}\sum_{i=1}^{n_F}\sum_{k=1}^{K}
(\alpha_k+\gamma+\upsilon_k) s_{i,k} \tilde{C}_{i,k}\hspace*{-0.5mm}-\hspace*{-0.5mm}\notag\lambda\Big(\sum_{i=1}^{n_F}\sum_{k=1}^{K}\tilde{P}_{i,k}-P_{\max} \Big)\notag\\ \notag\notag\\
&&\hspace*{-4mm}\notag-q\Big(P_{C_T}+ K P_{C_R}+
\sum_{k=1}^{K} \sum_{i=1}^{n_F}\varepsilon \tilde{P}_{i,k} \Big.\\
  \Big.\hspace*{-4mm}&-&\hspace*{-4mm}\sum_{k=1}^{K}\sum_{i=1}^{n_F} \Big(\sum_{j=1}^{K} \tilde{P}_{i,j}\Big) l_k g_k \abs{H_{i,k}}^2 \eta_k- \sum_{k=1}^{K}Q_{I_k}\Big)-\gamma R_{\min}\notag\\
  \hspace*{-4mm}&-&\hspace*{-4mm}\beta\Big(P_{PG}- P_{C_T}+\sum_{i=1}^{n_F}\sum_{k=1}^{K}\varepsilon \tilde{P}_{i,k} \Big)\notag\\
\hspace*{-4mm}&-&\hspace*{-4mm}\sum_{i=1}^{n_F}\sum_{k=1}^K\delta_{i,k}\Big(\tilde{\rho}^I_{i,k} + \rho^E_{i,k}-1\Big)-\sum_{i=1}^{n_F}\zeta_i \Big(\sum_{k=1}^K s_{i,k}-1\Big)\notag\\
\hspace*{-50mm}&-&\notag\hspace*{-4mm}  \sum_{k=1}^K \sum_{i=2}^{n_F}\mu_{i,k} \Big({\rho}^E_{1,k}-{\rho}^E_{i,k}\Big)-\sum_{k\in{\cal D}}\upsilon_k R_{\min_k}\\
\hspace*{-55mm}&-&\hspace*{-4mm} \sum_{k=1}^K w_k\Big( \frac{P_{\min_k}^{\rm req}}{\rho_{1,k}^E}\hspace*{-1mm}- \hspace*{-1mm} \sum_{i=1}^{n_F}\hspace*{-1mm} \Big(\hspace*{-1mm}\sum_{j=1}^{K}\tilde{P}_{i,j}\hspace*{-1mm}\Big) l_k g_k \abs{H_{i,k}}^2 \eta_k\hspace*{-1mm}-\hspace*{-1mm} \sum_{i=1}^{n_F}\hspace*{-0.5mm}\sigma_{I_{i,k}}^2  \eta_k \Big),\notag
\label{eqn:Lagrangian}
\end{eqnarray}
where  $\tilde{C}_{i,k}=W\log_2\Big(\frac{\tilde{P}_{i,k}}{s_{i,k}}\tilde{\Gamma}_{i,k}\Big)$ and $\tilde{\Gamma}_{i,k}=\Gamma_{i,k}\Big|_{{\rho}^I_{i,k}=\frac{\tilde{\rho}^I_{i,k}}{s_{i,k}}}$.
$\mathbf{w}$ has elements $w_k\ge0$, $k\in\{1,\ldots,K\}$, and is the Lagrange multiplier vector  corresponding to the individual minimum required  power transfer constraint. $\lambda\ge0$
accounts for the maximum transmit power allowance $P_{\max}$ in constraint C2. $\beta\ge0$ is the Lagrange multiplier for C3 accounting for the power usage from
the power grid at the transmitter. $\gamma$ is the Lagrange multiplier associated with the minimum data rate requirement of the system in C4.  $\mbox{\boldmath$\upsilon$}$ denotes the Lagrange multiplier vector for the minimum individual data rate requirement of receiver $k$ in C5 and has elements $\upsilon_k\ge0, k\in\{1,\ldots,K\}$. We note that  $\upsilon_k =0,  \forall k\notin {\cal D}$, for the receivers requiring non-delay constrained services. $\mbox{\boldmath$\zeta$}$ is the Lagrange multiplier vector accounting for the subcarrier assignment constraint C8 and has elements $\zeta_i, i\in\{1,\ldots,n_F\}$. $\mbox{\boldmath$\delta$}$ is the Lagrange multiplier vector for the power splitting constraint C11 and has elements $\delta_{i,k}, i\in\{1,\ldots,n_F\},k\in\{1,\ldots,K\}$. $\mbox{\boldmath$\mu$}$ is the Lagrange multiplier vector  for constraint C12 and has elements $\mu_{i,k},i\in\{1,\ldots,n_F\},k\in\{1,\ldots,K\}$.
  On the other hand, the boundary constraints C6, C7, C9, and C10 on the optimization variables  are captured by the Karush-Kuhn-Tucker (KKT) conditions
when deriving the resource allocation solution in the next section.
Thus, the dual problem for the primal problem (\ref{eqn:inner_loop}) is given by
\begin{eqnarray}
\underset{ \mathbf{w}, \lambda,\gamma,\beta,{\boldsymbol\delta},
\boldsymbol{\upsilon},\boldsymbol{\zeta}\ge 0,{\boldsymbol\mu}}{\min}\ \underset{{\cal
P,S, {\boldsymbol\rho}}}{\max}\quad{\cal L}( \mathbf{w}, \lambda,\gamma,\beta,\mbox{\boldmath$\delta$},\boldsymbol{\upsilon},\mbox{\boldmath$\mu$},\mbox{\boldmath$\zeta$},{\cal P},{\cal S},{\mbox{\boldmath$\rho$}}).\label{eqn:master_problem}
\end{eqnarray}

\subsection{Dual Decomposition Solution }
\label{sect:sub_problem_solution}
In this section, the optimal\footnote{In this section, an optimality  refers to the optimality for the problem formulation using the time-sharing assumption and the approximated objective function.  } resource allocation policy is obtained via Lagrange dual decomposition \cite{CN:large_subcarriers}. Specifically, the dual
problem in (\ref{eqn:master_problem}) is decomposed into a hierarchy of two levels. Level 1, the inner maximization in (\ref{eqn:master_problem}), consists of $n_F$ subproblems with identical structure that can be solved in parallel.  Level 2,  the outer minimization in (\ref{eqn:master_problem}),  is the master problem. The dual problem can be
solved iteratively. Specifically,   in each iteration,  the transmitter solves the
 $n_F$ subproblems  by applying the KKT conditions for a fixed set of Lagrange multipliers. Then the solutions of the subproblems are used  for updating the Lagrange multiplier  master problem via the gradient method.

\subsubsection*{Level 1 (Subproblem Solution)}

 Using standard convex
optimization techniques and the KKT conditions, for a given  $q$, in each iteration of the Dinkelbach method,    the power allocation policy and the power splitting policy
on subcarrier $i$ for receiver $k$ are given by
 \begin{eqnarray}\label{eqn:power1}
\hspace*{-5mm}\tilde{P}_{i,k}^*\hspace*{-2mm}&=&\hspace*{-2mm}s_{i,k}{P}_{i,k}^*=s_{i,k}
\Bigg[\frac{W(\alpha_k+\gamma+\upsilon_k)}
{\ln(2)\Big(\Phi_{i,k}\Big)}\Bigg]^+\hspace*{-3mm}, \,\forall i,k,\,\end{eqnarray}

 \begin{eqnarray}  \hspace*{-5mm} \Phi_{i,k}\hspace*{-3mm}&=&\hspace*{-3mm} q\varepsilon +\beta\varepsilon+\lambda - \sum_{k=1}^K(q+w_k)l_k g_k \abs{H_{i,k}}^2 \eta_k,\\
\label{eqn:opt_rho_E}
\hspace*{-5mm}\rho^{E*}_{1,k}\hspace*{-2mm}&=&\hspace*{-2mm} \Bigg[ \sqrt{\Big[\frac{w_k P_{\min_k}^{\rm req}}{\delta_{1,k}-q(\sigma_{z^a}^2+\sigma_{I_{1,k}}^2) \eta_k+\sum_{j=2}^{n_F}\mu_{j,k}}\Big]^+}\Bigg]_{\rho^E_{L}}^{\rho^E_{U}}\hspace*{-2mm},\\
\notag&&\forall k,\,\mbox{and}\\
\label{eqn:opt_rho_E2}\rho^{E*}_{j,k}\hspace*{-2mm}&=&\hspace*{-2mm} \Bigg[q(\sigma_{z^a}^2+\sigma_{I_{j,k}}^2)\eta_k - \delta_{j,k}+\mu_{j,k} \Bigg]_{\rho^E_{L}}^{\rho^E_{U}},\,\\
\notag &&\forall k,j\in\{2,\ldots,n_F\},\\
\label{eqn:opt_rho_I}
\hspace*{-5mm}\tilde{\rho}^{I*}_{i,k}\hspace*{-2mm}&=&\hspace*{-2mm} s_{i,k}{\rho}^{I*}_{i,k}\hspace*{-1mm}=\hspace*{-1mm}s_{i,k}\Bigg[\frac{\Psi_{i,k}}{2\sqrt{\ln(2)(\delta_{i,k})}
(\sigma_{z^a}^2+\sigma_{I_{i,k}}^2)}\notag\\
&&-\frac{\sigma_{z^s}^2}{2(\sigma_{z^a}^2+\sigma_{I_{i,k}}^2)}\Bigg]_{\rho^I_{L}}^{\rho^I_{U}},\hspace*{-1mm}\forall i,k,\,\\
   \hspace*{-5mm}\Psi_{i,k}\hspace*{-3mm}&=&\hspace*{-3mm}
    \sqrt{4 W(\alpha_k+\gamma+\upsilon_k)
   (\sigma_{z^a}^2+\sigma_{I_{i,k}}^2)+ (\delta_{i,k}) \ln (2) \sigma_{z^s}^2} \quad \notag\\
  &&\hspace*{-3mm} \times\sqrt{\sigma_{z^s}^2} \label{eqn:phi}
\end{eqnarray}
 and $\mu_{1,k}=0$. Here, operators $\big[x\big]^+$ and $\big[x\big]^c_d$ are defined as $\big[x\big]^+=\max\{0,x\}$ and  $\big[x\big]^c_d=c,\ \mbox{if}\ x>c,\big[x\big]^c_d=x,\mbox{ if}\, d\le x\le c,\big[x\big]^c_d  =d,\ \mbox{if}\ d>x$, respectively. The
    power allocation solution in (\ref{eqn:power1}) is known as  multilevel water-filling. In particular, the water-level in allocating power on subcarrier $i$ for receiver $k$, i.e., $\frac{W(\alpha_k+\gamma+\upsilon_k)}{\ln(2)\Phi_{i,k}}$, is not only directly proportional to  the priority  of receiver $k$ via variable  $\alpha_k$, but also  depends on the channel gains of the other $K-1$ receivers via the term $\sum_{k=1}^K(q+w_k)l_k g_k \abs{H_{i,k}}^2 \eta_k$. Besides, Lagrange multipliers $\gamma$, $\upsilon_k$, and $w_k$
force the transmitter to transmit with a sufficiently high power to fulfill the system data rate requirement,
 $R_{\min}$, the individual data rate requirements of the receivers having  delay
 constrained services, $R_{\min_k}$,  and the minimum power transfer requirement, $P_{\min_k}^{\rm req}$, for receiver $k$, respectively. Moreover, as can be observed from (\ref{eqn:power1}), the power allocation solution $P_{i,k}^*$ is independent of $s_{i,k}$ which facilitates a simple allocation design.

  On the other hand, the power splitting ratio for information decoding, ${\rho}^{I*}_{i,k}$, is also in the form of water-filling and the water-level depends on the priority of the receiver via $\alpha_k$ in  (\ref{eqn:opt_rho_I}). Besides, Lagrange multiplier $\mu_{i,k}$ affects the power splitting ratio solution for energy harvesting in   (\ref{eqn:opt_rho_E}) and (\ref{eqn:opt_rho_E2}) such that $\rho^E_{1,k} $ will eventually equal $\rho^E_{j,k},\forall j\in\{2,\ldots,n_F\}$, as enforced  by consensus constraint C12. Furthermore, when $\sigma_{z^a}^2+\sigma_{I_{i,k}}^2\gg \sigma_{z^s}^2$, the SINR  on each subcarrier approaches $\frac{P_{i,k}\rho_{i,k}^{I*} l_k
g_k\abs{H_{i,k}}^2}{\rho_{i,k}^{I*}(\sigma_{z^a}^2+\sigma_{I_{i,k}}^2)+\sigma_{z^s}^2}\rightarrow \frac{P_{i,k} l_k
g_k\abs{H_{i,k}}^2}{\sigma_{z^a}^2+\sigma_{I_{i,k}}^2} $ and is independent of ${\rho}^{I*}_{i,k}$. Besides,  $ \widehat {U}_{TP}({\cal P},{\cal S},\mbox{\boldmath$\rho$})$ is a monotonically decreasing function of ${\rho}^{E*}_{i,k}$. Therefore, ${\rho}^{E*}_{i,k}\rightarrow {\rho}^{E}_{U}$ and ${\rho}^{I*}_{i,k}\rightarrow {\rho}^{I}_{L}$ become the optimal power splitting policy for energy efficiency maximization. Thus, if $\sigma_{z^a}^2+\sigma_{I_{i,k}}^2\gg \sigma_{z^s}^2$ and the lower bound of the power splitting ratio for information decoding is zero, i.e.,   ${\rho}^{I}_{L}=0$, then the solution suggests that in this case an infinitesimally small  portion of the received power is used at the receivers  for information decoding in order to achieve the optimal performance; provided that the data rate constraints C4 and C5 are satisfied. In other words, most of the received power at the receivers should be used for energy harvesting. The above considerations indicate that in the interference limited regime, a hybrid information and energy  harvesting receiver must achieve a higher energy efficiency than the traditional pure information receiver. This will be confirmed by simulation in Section V.

 On the other hand, subcarrier $i$ is assigned to receiver $k$ when the following selection criterion is satisfied:  \begin{eqnarray}
\label{eqn:sub_selection} s_{i,k}^*\hspace*{-3mm}&=&\hspace*{-3mm}
 \left\{ \begin{array}{rl}
 1 &\mbox{if $k=\arg \underset{a}{\max} \quad  \ M_{i,a}$ }, \\
 0 &\mbox{ otherwise}
       \end{array} \right.  \\  \label{eqn:marginal}
  \mbox{where   }     &&M_{i,k}\\
  &=&W\Big (\alpha_k+\gamma+\upsilon_k\Big)\Big[\log_2\Big(P_{i,k}^* l_k
g_k\abs{H_{i,k}}^2\Big)\notag\\
&&+\log_2\Big(\frac{\rho^{I*}_{i,k}}
{(\rho^{I*}_{i,k})(\sigma_{z^a}^2+\sigma_{I_{i,k}}^2)+\sigma_{z^s}^2}\Big)\notag\\
&&-\frac{1}{\ln(2)}-\frac{\sigma_{z^s}^2}{\ln(2)(\rho^{I*}_{i,k}(\sigma_{z^a}^2+\sigma_{I_{i,k}}^2)+\sigma_{z^s}^2)}\Big]-\zeta_i
\notag\end{eqnarray}
is the marginal benefit provided to the system when subcarrier $i$ is assigned to serve receiver $k$. In other words, receiver $k$ is  selected for information transmission on subcarrier $i$ if it can provide the maximum marginal benefit to the system. Besides, if receiver $k$ has a high priority or a stringent individual data rate requirement, it will have  high values of $\alpha_k$ or $\upsilon_k$ and the resource allocator at the transmitter will have a higher preference to serve receiver $k$ with subcarrier $i$. On the other hand, it can be observed from (\ref{eqn:sub_selection}) that although constraint relaxation  is used in constraint C7 for facilitating   the design of the resource allocation algorithm,  the  subcarrier allocation policy on each subcarrier  for the relaxed problem remains  Boolean; time sharing does not occur.

\begin{Remark}\emph{The above observation for the optimal power splitting policy in the interference limited regime is valid for both the problem with the original objective function and the problem with the approximated objective function. Indeed, if the case of an interference limited regime is considered, the use of the approximated objective function is not necessary. Instead, we can first set ${\rho}^{E*}_{i,k}\rightarrow {\rho}^{E}_{U}$ and ${\rho}^{I*}_{i,k}\rightarrow {\rho}^{I}_{L}$ in the problem formulation.  ${\rho}^{E*}_{i,k}$ and ${\rho}^{I*}_{i,k}$  become constants and the associated non-convexity vanishes. Then,  we  optimize $\cal P $ and $\cal S$ by following  a similar approach as used in  (\ref{eqn:cross-layer})-(\ref{eqn:marginal}).}
\end{Remark}

\subsubsection*{Level 2 (Master Problem Solution)}
The Level 2 master problem in
(\ref{eqn:master_problem}) can be solved by using the gradient method which leads to the following Lagrange multiplier update equations:
\begin{eqnarray}\label{eqn:multipler1}
\hspace*{-5.5mm}\lambda(u+1)\hspace*{-3mm}&=&\hspace*{-3mm}\Big[\lambda(u)-\xi_1(u)\hspace*{-1mm}\times\hspace*{-1mm}
\Big(  P_{\max}-\sum_{i=1}^{n_F}\sum_{k=1}^{K}\tilde{P}_{i,k}\Big)\Big]^+\hspace*{-3mm},\\
\hspace*{-5.5mm}\beta(u+1)\hspace*{-3mm}&=&\hspace*{-3mm}\Big[\beta(u)-\xi_2(u)\times
\Big( P_{PG} -P_C\\ \notag
&&-\sum_{i=1}^{n_F}\sum_{k=1}^{K}\varepsilon \tilde{P}_{i,k})\Big]^+\hspace*{-1.5mm}, \label{eqn:multipler2}\\
\hspace*{-5.5mm}\gamma(u+1)\hspace*{-3mm}&=&\hspace*{-3mm}\Big[\gamma(u)-\xi_3(u)\times
\Big(\sum_{i=1}^{n_F}\sum_{k=1}^{K} s_{i,k} \tilde{C}_{i,k}\notag\\
&&-R_{\min}\Big )\Big]^+\hspace*{-1.5mm},\label{eqn:multipler3}\\ \notag
\hspace*{-5.5mm}\delta_{i,k}(u+1)\hspace*{-3mm}&=&\hspace*{-3mm}\Big[\delta_{i,k}(u)-\xi_4(u)\\
&&\times
\Big(1-\tilde{\rho}^I_{i,k} - \rho^E_{i,k}\Big)\Big]^+,\forall i, k, \label{eqn:multipler4}\\
\hspace*{-5.5mm}\mu_{r,k}(u+1)\hspace*{-3mm}&=&\hspace*{-3mm}\Big[\mu_{r,k}(u)-\xi_5(u)\times
\Big({\rho}^E_{r,k}-{\rho}^E_{1,k}\Big)\Big],\\
&&\notag\forall r\in\{2,\ldots,n_F\}, \forall k, \label{eqn:multipler5}\\
\hspace*{-5.5mm}w_{k}(u+1)\hspace*{-3mm}&=&\hspace*{-3mm}\Big[w_{k}(u)-\xi_6(u)\Big.\notag\\
\hspace*{-3mm}&&\times\Big.
\Big(\sum_{i=1}^{n_F} \Big(\sum_{j=1}^{K}\tilde{P}_{i,j}\Big) l_k g_k \abs{H_{i,k}}^2 \eta_k\notag\\
\hspace*{-3mm}&& + \sum_{i=1}^{n_F}\sigma_{I_{i,k}}^2  \eta_k - \frac{P_{\min_k}^{\rm req}}{\rho_{1,k}^E}, \Big)\Big]^+,\forall  k, \label{eqn:multipler7}\\
\upsilon_k(u+1)\hspace*{-3mm}&=&\hspace*{-3mm}\Big[\upsilon_k(u)-\xi_7(u)\times
\Big(\sum_{i=1}^{n_F}s_{i,k}\tilde{C}_{i,k}\notag\\
\hspace*{-3mm}&& -R_{\min_k}\Big )\Big]^+, \forall k\in{\cal D},\label{eqn:multipler8}
\end{eqnarray}
where index $u\ge 0$ is the iteration index and $\xi_t(m)$,
$t\in\{1,\ldots,7\}$, are positive step sizes.  Then, the updated\footnote{It can be observed that updating $\zeta_i$ is not necessary since it does not affect the result of the subcarrier allocation solution in (\ref{eqn:sub_selection}). } Lagrange multipliers in
(\ref{eqn:multipler1})--(\ref{eqn:multipler8}) can be used for updating the resource allocation policy
 in (\ref{eqn:power1})--(\ref{eqn:marginal}) by solving
the $n_F$ subproblems in (\ref{eqn:master_problem}). As the primal problem with the approximated objective function  is jointly concave w.r.t. the optimization variables, it is
guaranteed that the primal optimal solution can be obtained by solving the problems in Level 1 and Level 2 iteratively, provided that the chosen step
 sizes, $\xi_t(m)$, are sufficiently small\footnote{In the literature, different methods such as exact line search and backtracking line search are commonly used for optimizing the step sizes \cite{book:convex,Notes:Sub_gradient}. In this paper, we adopt a backtracking line search approach in optimizing the step sizes for fast convergence. }. We note that  other than the adopted  gradient method, different iterative algorithms such as the ellipsoid method  can also be used for finding the optimal Lagrange multipliers due to the convexity of the dual problem \cite{book:convex}.

\section{Resource Allocation Design - Discrete Set of Power Splitting Ratios}
 \label{sect4}
 In practice, due to the high complexity associated with a high precision power splitting unit, the RF energy harvesting receivers may only be capable of splitting the received power into two power streams based on a finite discrete set of power splitting ratios. In this section, we design a resource allocation algorithm  for  such receivers. In particular, we assume that there are $N$ distinct power splitting ratios for  energy harvesting and information decoding at each receiver. Thus, the power splitting ratios for energy harvesting  and information decoding   on subcarrier $i$ for mobile receiver $k$ can be represented by the following constraints:
\begin{eqnarray}
\mbox{C14:}\,\,\rho^{E}_{i,k}&\in&\{\rho^{E_1}_{k},\rho^{E_2}_{k},\ldots,
\rho^{E_n}_{k},\ldots,\rho^{E_N}_{k}\},\notag\\
\mbox{C15:}\,\,\rho^{I}_{i,k}&\in&\{\rho^{I_1}_{k},
\rho^{I_2}_{k},\ldots,\rho^{I_n}_{k},\ldots, \rho^{I_N}_{k}\}. \label{eqn:rho_E-discrete}
 \end{eqnarray}
 $\rho^{E_n}_{k}$ and $\rho^{I_n}_{k}$, $n\in\{1,2,\ldots,N\}$, are the possible power splitting modes  for energy harvesting  and information decoding adopted  in receiver $k$, respectively. The corresponding resource allocation algorithm design can be formulated as the following optimization problem:
\begin{eqnarray}
\label{eqn:cross-layer-discrete-original}&&\hspace*{5mm} \max_{{\cal P},{\cal S}, {\boldsymbol \rho}}\,\, U_{eff}({\cal P},{\cal S},\mbox{\boldmath$\rho$})  \nonumber\\
\hspace*{-5mm}\mbox{s.t.} &&\mbox{C1--C8, C11, C12, C14, C15}.
\end{eqnarray}
Similar to the case of a continuous set of power splitting ratios, the objective function of the above problem formulation inherits the non-convexity of the fractional form of the objective function. Therefore, by using Theorem 1, we can transform the objective function from the fractional form into a subtractive form and solve the problem via the Dinkelbach method.  As shown in Table \ref{table:algorithm}, in each iteration of the main loop, i.e., lines 3--12,  we solve the following optimization problem for a given
parameter $q$:
\begin{eqnarray}
\label{eqn:cross-layer-discrete-original}&&\hspace*{-5mm} \max_{{\cal P},{\cal S}, {\boldsymbol \rho}} \,U({\cal
P},{\cal S}, {\boldsymbol \rho})-q U_{TP}({\cal P},{\cal S},{\boldsymbol \rho})  \nonumber\\
\hspace*{-5mm}\mbox{s.t.} &&\mbox{C1--C8, C11, C12, C14, C15}.
\end{eqnarray}

The additional difficulty in solving the above optimization problem compared to the problem formulation in (\ref{eqn:cross-layer}) is  the disjoint/discrete nature of the optimization variables $\rho^{E}_{i,k}$ and $\rho^{I}_{i,k}$, cf. C14 and C15 in (\ref{eqn:rho_E-discrete}). In general, an exhaustive search is required to obtain the global optimal solution and the search space grows in the order of $N^{2K^{n_F}}$, which may not be computationally feasible for systems of moderate size. In the following, we transform (\ref{eqn:cross-layer-discrete-original}) into an optimization problem with tractable solution by exploiting subcarrier time-sharing and the concept of power splitting mode selection. In particular,  different values of  $\rho^{I_n}_{k}$ can be treated as different operating modes of receiver $k$ with different equivalent SINRs. Then, we combine the subcarrier selection with the operating mode selection by augmenting the dimensions of the optimization variables.  To this end, we define the channel capacity between the transmitter  and  receiver $k$ on subcarrier $i$ with channel bandwidth $W$ by using power splitting mode $n$ as
\begin{eqnarray}\label{eqn:cap-operating_mode}
C_{i,k}^n&=&W\log_2\Big(1+P_{i,k}^n\Gamma_{i,k}^n\Big)\\
\notag\,\,\Gamma_{i,k}^n&=&\frac{\rho_{k}^{I_n} l_k
g_k\abs{H_{i,k}}^2}{\rho_{k}^{I_n}(\sigma_{z^a}^2+\sigma_{I_{i,k}}^2)+\sigma_{z^s}^2},
\end{eqnarray}
and $P_{i,k}^n \Gamma_{i,k}^n$ is the received SINR on subcarrier $i$ at receiver $k$ using power splitting mode $n$ for information decoding. The \emph{weighted system capacity} is defined as the total
average number of  bits successfully delivered to the $K$ receivers via the $N$ power splitting modes  and is given by
\begin{eqnarray}
 \label{eqn:discrete} && \hspace*{-5mm} U({\cal P}^N,{\cal S}^N)=\sum_{i=1}^{n_F} \sum_{k=1}^{K}\sum_{n=1}^{N} \alpha_k s_{i,k}^n C_{i,k}^n,
\end{eqnarray}
where  ${\cal P}^N=\{ P_{i,k}^n \ge 0, \forall i,k,n\}$ is the power allocation policy and ${\cal S}^N=\{ s_{i,k}^n\in\{0,1\}, \forall i,k,n\}$ is the subcarrier allocation policy for the case of discrete power splitting ratios. We note that the subcarrier allocation policy in (\ref{eqn:discrete}) incorporates the power splitting mode selection for information decoding. On the other hand, the power consumption of the system can be written as
\begin{eqnarray}
&&U_{TP}({\cal P}^N,{\cal S}^N,{\cal A}^N)\notag\\
&=&P_{C_T}+ K P_{C_R}+
 \sum_{i=1}^{n_F}\sum_{k=1}^{K}\sum_{n=1}^{N}\varepsilon P_{i,k}^ns_{i,k}^n \notag\\
 &&\hspace*{-5mm}-  \sum_{k=1}^{K}\sum_{n=1}^{N}Q_{D_k}^n - \sum_{k=1}^{K}\sum_{n=1}^{N}Q_{I_k}^n \\
\hspace*{-5mm}\mbox{where}\,Q_{D_k}^n \hspace*{-4mm}&=& \hspace*{-4mm}\underbrace{\sum_{i=1}^{n_F} \hspace*{-1mm} \Big(\sum_{j=1}^{K}\sum_{m=1}^{N}\hspace*{-1mm}P_{i,j}^m s_{i,j}^m\Big) l_k g_k \abs{H_{i,k}}^2 \eta_k \rho_{k}^{E_n}a_{k}^n }_{\underset{\mbox{at receiver $k$ with power splitting mode $n$}}{\mbox{Power harvested  from information signal  }}}\\ \mbox{and}\,\,\,\, Q_{I_k}^n \hspace*{-4mm}&=& \hspace*{-4mm} \hspace*{-20mm}\underbrace{\sum_{i=1}^{n_F}(\sigma_{z^a}^2+\sigma_{I_{i,k}}^2 ) \rho_{k}^{E_n}a_{k}^n \eta_k}_{\underset{\mbox{antenna noise at receiver $k$ with power splitting mode $n$}}{\mbox{Power harvested  from interference and }}}\hspace*{-7mm}.
\end{eqnarray}\newline
Here, ${\cal A}^N=\{ a_{k}^n\in\{0,1\}, \forall k,n\}$ is the power splitting ratio selection policy for energy harvesting. In the above problem formulation, $a_{k}^n$ is the optimization variable which  captures the selection of power splitting mode $\rho_{k}^{E_n}$ for energy harvesting.  Similar to the case of the continuous set of power splitting ratios, we consider an approximation of the objective function for facilitating a tractable resource allocation algorithm design in the following. First, the system capacity between the transmitter and the $K$ mobile receivers can be approximated by
\begin{eqnarray}\notag \label{eqn:discrete-approx}\notag
\hspace*{-5mm}{U}({\cal P}^N,{\cal S}^N ) &\stackrel{(a)}{\approx}& \widehat{U}({\cal P}^N,{\cal S}^N )\\ \notag \hspace*{-1mm}&=&\hspace*{-1mm}\sum_{i=1}^{n_F} \sum_{k=1}^{K}\sum_{n=1}^{N} \alpha_k s_{i,k}^n \widehat{C}_{i,k}^n,\\
\widehat{C}_{i,k}^n\hspace*{-1mm}&=&\hspace*{-1mm}W\log_2\Big(P_{i,k}^n\Gamma_{i,k}^n\Big) \end{eqnarray}
and $\stackrel{(a)}{\approx} $ in (\ref{eqn:discrete-approx}) is due to the high SINR assumption.
On the other hand, a lower bound for the total power consumption of the system is given by
\begin{eqnarray}
&&U_{TP}({\cal P}^N,{\cal S}^N,{\cal A}^N)\notag\\
\hspace*{-1mm}&\ge&\hspace*{-1mm} \widehat {U}_{TP}({\cal P}^N,{\cal S}^N,{\cal A}^N)\notag\\
&=&P_{C_T}+ K P_{C_R}+  \sum_{k=1}^{K}\sum_{i=1}^{n_F}\sum_{n=1}^{N}\varepsilon P_{i,k}^n s_{i,k}^n
  \notag\\
 \hspace*{-1mm}&-&\hspace*{-3mm} \sum_{i=1}^{n_F}\sum_{k=1}^{K} \Big(\sum_{j=1}^{K}\sum_{m=1}^{N}P_{i,j}^m s_{i,j}^m\Big) l_k g_k \abs{H_{i,k}}^2\eta_k \notag\\
 &- &\hspace*{-1mm} \sum_{k=1}^{K}\sum_{n=1}^{N}\sum_{i=1}^{n_F}Q_{I_k}^n,
\end{eqnarray}
where $ \widehat {U}_{TP}({\cal P}^N,{\cal S}^N,{\cal A}^N)$ is obtained by setting $\rho_{k}^{E_n}a_{k}^n=1$. Then, the  objective function for the optimization problem for discrete sets of power splitting factors  is given by
\begin{eqnarray}
&&{U}({\cal P}^N,{\cal S}^N)-q{U}_{TP}({\cal P}^N,{\cal S}^N, {\cal A}^N
)\notag\\
&\lesssim&  \widehat{U}({\cal P}^N,{\cal S}^N )-q \widehat {U}_{TP}({\cal P}^N,{\cal S}^N,{\cal A}^N).
\end{eqnarray}
\subsection{Optimization Problem Formulation}
\label{sect:cross-Layer_formulation-discrete}
With a slight abuse of notation, we reformulate the optimization problem for  discrete power splitting ratios  to be solved in each iteration as follows:
\begin{eqnarray}
\label{eqn:cross-layer-discrete}&&\hspace*{-5mm} \max_{{\cal P}^N,{\cal S}^N, {\cal A}^N}\,\, \widehat{U}({\cal P}^N,{\cal S}^N )-q \widehat {U}_{TP}({\cal P}^N,{\cal S}^N,{\cal A}^N)\nonumber\\
\notag \hspace*{-2mm}\mbox{s.t.} &&\notag\hspace*{-5mm}\mbox{C1: }
\sum_{n=1}^N Q_{D_k}^n+Q_{I_k}^n\ge P_{\min_k}^{\rm req},\forall k,\notag\\
&& \hspace*{-5mm}\mbox{C2:}\notag\sum_{i=1}^{n_F}\sum_{k=1}^{K}\sum_{n=1}^{N}P_{i,k}^n s_{i,k}^n \le P_{\max},\\ &&\hspace*{-5mm} \mbox{C3:}\notag\,\,P_C+\sum_{i=1}^{n_F}\sum_{k=1}^{K}\sum_{n=1}^{N}\varepsilon P_{i,k}^n s_{i,k}^n \le P_{PG},\notag \\
&&\hspace*{-5mm}\notag  \mbox{C4:}\sum_{i=1}^{n_F}\sum_{k=1}^{K}\sum_{n=1}^{N}s_{i,k}^n \widehat{C}_{i,k}^n\ge R_{\min} ,\\ &&\hspace*{-5mm}\notag\mbox{C5:}\,\, \sum_{i=1}^{n_F}\sum_{k=1}^{K}\sum_{n=1}^{N}s_{i,k}^n \widehat{C}_{i,k}^n\ge R_{\min_k}, \forall k\in {\cal D},\\
&&\hspace*{-5mm}\notag\mbox{C6:}\,\, P_{i,k}^n\ge 0, \,\, \forall i,k,n,\quad\notag\,\,\, \mbox{C7:}\,\, s_{i,k}^n\in\{0,1\},\forall i,k,n,\\
&&\hspace*{-5mm}\mbox{C8:}\notag\,\, \sum_{k=1}^{K}\sum_{n=1}^{N} s_{i,k}^n\le 1, \,\, \forall i, \,\mbox{C9:}\notag\,\,  a_{k}^n\in\{0,1\}, \,\, \forall k,n,
\end{eqnarray}\begin{eqnarray}
&&\hspace*{-7mm}\notag \mbox{C10:}\notag\,\,  \sum_{n=1}^N a_{k}^n=1 ,\forall k,\\&&\hspace*{-7mm} \notag\mbox{C11:}\,\, \rho^{I_n}_{k} s_{i,k}^n+ \sum_{m=1}^{N}\rho^{E_m}_{k}a_{k}^m\le 1, \,\, \forall i,k,n,\\
&&\hspace*{-7mm}\mbox{C12:}\, \sum_{m=1}^{N}\rho^{I_m}_{k} s_{i,k}^m+ \rho^{E_n}_{k}a_{k}^n\le 1, \,\, \forall i,k,n.
\end{eqnarray}
Constraints C1--C6 have the same physical meanings  as in the problem formulation in (\ref{eqn:cross-layer}).
Constraints C7--C12 are imposed to guarantee that in each receiver only one power splitting mode can be selected  for  information decoding and energy harvesting. Besides, C11 and C12 indicate that no extra power gain can be achieved in the power splitting process.  On the other hand, the non-convexity of the above problem formulation  is caused by the coupled optimization variables in C1 and  the combinatorial constraints in C7 and C9. In analogy to the techniques used for solving the optimization problem in (\ref{eqn:inner_loop}), we solve problem (\ref{eqn:cross-layer-discrete}) in the following two steps. In the first step, we relax constraints C7 and C9 such that variables $s_{i,k}^n$ and $a_{i,k}^n $ can assume any value between zero and one, i.e., $\mbox{C7:}\,\, 0\le s_{i,k}^n\le 1,\forall i,k,n$ and $\mbox{C9:}\,\, 0\le a_{i,k}^n\le 1,\forall i,k,n$. Then $s_{i,k}^n$ and $a_{i,k}^n$ can be interpreted  as the time sharing factors for receiver $k$ in utilizing subcarrier $i$ with power splitting mode $n$. We also define a new variable $\tilde{P}_{i,k}^n=s_{i,k}^n{P}_{i,k}^n$   for facilitating  the design of the resource allocation algorithm. In fact, $\tilde{P}_{i,k}^n$ represents the actual transmit power of the transmitter for  receiver $k$ in subcarrier $i$ if power splitting mode $n$ is used under the time sharing condition. In the second step, we replace constraint C1 in (\ref{eqn:cross-layer-discrete}) by
\begin{eqnarray}\label{eqn:constraint_c1_discrete}
\mbox{C1': }  &&\sum_{i=1}^{n_F} \Big(\sum_{j=1}^{K}\sum_{m=1}^{N}\tilde{P}_{i,j}^m \Big) l_k g_k \abs{H_{i,k}}^2 \eta_k \notag\\
 &&+\sum_{i=1}^{n_F}(\sigma_{z^a}^2+\sigma_{I_{i,k}}^2 ) \eta_k \ge \frac{a_{k}^n P_{\min_k}^{\rm req}}{\rho_{k}^{E_n}} , \forall k,n.
\end{eqnarray}
Although C1' is  equivalent to C1 in (\ref{eqn:cross-layer-discrete}) only if $a_{k}^n $ takes
a binary value, i.e., $a_{k}^n \in\{0,1\}$, and  $\sum_{k=1}^Ka_{k}^n =1$, it will be shown in the next section that the optimal solution for $a_{k}^n$ has a binary form under constraint C1', despite the adopted constraint relaxation. Consequently, the optimization problem with the approximated objective function and the constraint relaxation is now jointly concave w.r.t. all optimization variables\footnote{The concavity of the above optimization problem can be proved by following a similar approach as in Appendix B for the case of continuous power splitting ratios.  }. Besides, it satisfies Slater's constraint qualification. Therefore, we can apply dual decomposition
 to solve the primal problem via solving its dual problem. The Lagrangian function of the primal problem in (\ref{eqn:cross-layer-discrete}) is given by
\begin{eqnarray}&&{\cal
L}( \mathbf{w}, \lambda,\gamma,\beta,\mbox{\boldmath$\delta$},\boldsymbol{\upsilon},
\mbox{\boldmath$\varphi$},\boldsymbol{\kappa},\boldsymbol{\zeta},{\cal P}^N,{\cal S}^N,{\cal A}^N)\\
\notag&=&\sum_{i=1}^{n_F}\sum_{k=1}^{K}\sum_{n=1}^{N}
(\alpha_k+\gamma+\upsilon_k) s_{i,k}^n \widehat{C}_{i,k}^n\end{eqnarray}\begin{eqnarray}
\notag&-&q\Big(P_{C_T}+ K P_{C_R}+
 \sum_{k=1}^{K}\sum_{n=1}^{N}\sum_{i=1}^{n_F}\varepsilon \tilde{P}_{i,k}^n \Big.\\
  \Big.\hspace*{-4mm}&-&\hspace*{-4mm}\sum_{k=1}^{K}\sum_{i=1}^{n_F}\sum_{n=1}^{N} \Big(\sum_{j=1}^{K} \tilde{P}_{i,j}^n\Big) l_k g_k \abs{H_{i,k}}^2 \eta_k- \sum_{k=1}^{K}\sum_{n=1}^{N}Q_{I_k}^n\Big)\notag\\
\hspace*{-4mm}&-&\hspace*{-3mm}\gamma R_{\min}-\sum_{i=1}^{n_F}\zeta_i \Big(\sum_{k=1}^K \sum_{n=1}^N s_{i,k}^n-1\Big)\notag\\
\hspace*{-4mm}&-&\hspace*{-4mm}\notag\lambda\Big(\sum_{i=1}^{n_F}
\sum_{k=1}^{K}\sum_{n=1}^{N}\tilde{P}_{i,k}^n-P_{\max} \Big)\notag\\
\hspace*{-4mm}&-&\sum_{i=1}^{n_F}\sum_{k=1}^K\sum_{n=1}^N\delta_{i,k}^n\Big(\rho^{I_n}_{k} s_{i,k}^n+ \sum_{m=1}^N\rho^{E_m}_{k}a_{k}^m- 1\Big)\notag\\
\hspace*{-4mm}&-&\hspace*{-4mm}\notag\sum_{i=1}^{n_F}\sum_{k=1}^K\sum_{n=1}^N\kappa_{i,k}^n
\Big(\sum_{m=1}^{N}\rho^{I_m}_{k} s_{i,k}^m+ \rho^{E_n}_{k}a_{k}^n-1\Big)\\
\hspace*{-4mm}&-&\hspace*{-4mm} \sum_{k=1}^K\sum_{n=1}^N w_{k}^n\Big(  \frac{a^n_k P_{\min_k}^{\rm req}}{\rho_{k}^{E_n}}-\sum_{i=1}^{n_F}\Big(  \sum_{j=1}^{K}\sum_{m=1}^{N}\tilde{P}_{i,j}^m \Big) l_k g_k \abs{H_{i,k}}^2 \eta_k\notag\\
\hspace*{-4mm}&-&\hspace*{-4mm}\sum_{i=1}^{n_F}(\sigma_{z^a}^2+\sigma_{I_{i,k}}^2 ) \eta_k \Big)-\sum_{k\in{\cal D}}\upsilon_k R_{\min_k}\notag\\
\hspace*{-4mm}&-&\hspace*{-4mm}\sum_{k=1}^{K} \varphi_{k}\Big(\hspace*{-1mm}\sum_{n=1}^N a_{k}^n-1\hspace*{-1mm}\Big)-\beta\Big(P_{PG}- P_{C_T}\hspace*{-1mm}+\hspace*{-1mm}\sum_{i=1}^{n_F}\sum_{k=1}^{K}\sum_{n=1}^{N} \varepsilon \tilde{P}_{i,k}^n \Big),\notag
\label{eqn:Lagrangian}
\end{eqnarray}where $\lambda$, $\beta$, and $\gamma$ are the scalar Lagrange multipliers associated to constraints C2, C3, and C4 in (\ref{eqn:cross-layer-discrete}), respectively. $\mathbf{w},\boldsymbol{\upsilon},
\boldsymbol{\varphi}$,  $\boldsymbol{\delta}$, and $\boldsymbol{\kappa}$ are the Lagrange multiplier vectors for constraints C1', C5, C8,  C10, C11, and C12
which have
elements $w_{k}^n\ge 0, n\in\{1,\ldots,N\},k\in\{1,\ldots,K\}$, ${\upsilon}_k\ge 0$, $\zeta_i\ge 0, \,i\in\{1,\ldots,n_F\}$, $\varphi_i$,
 $\delta_{i,k}^n$, and $\kappa_{i,k}^n$ respectively. We note that there is no restriction on the value of $\varphi_i$ since it is associated with equality constraint C10. Thus, the dual problem  is given by
\begin{eqnarray}
&&\underset{ \mathbf{w}, \lambda,\gamma,\beta,{\boldsymbol\delta},\boldsymbol{\upsilon},\boldsymbol{\zeta},\boldsymbol{\kappa}\ge 0, {\boldsymbol\varphi}}{\min}\\
 &&\underset{{\cal
P}^N,{\cal S}^N, {\cal A}^N}{\max}\quad{\cal
L}( \mathbf{w}, \lambda,\gamma,\beta,\mbox{\boldmath$\delta$},\boldsymbol{\upsilon}
,\mbox{\boldmath$\varphi$},\boldsymbol{\kappa},\boldsymbol{\zeta},{\cal P}^N,{\cal S}^N,{\cal A}^N).\label{eqn:master_problem-discrete}\notag
\end{eqnarray}

\subsection{Dual Decomposition Solution}
\label{sect:sub_problem_solution-discrete} By using dual decomposition and following a similar approach as in Section \ref{sect:sub_problem_solution},
 the resource allocation policy can be obtained via an iterative procedure. For a  given set of Lagrange multipliers $\{ \mathbf{w}, \lambda,\gamma,\beta,\mbox{\boldmath$\delta$},\boldsymbol{\upsilon},
 \mbox{\boldmath$\varphi$},\boldsymbol{\kappa},\boldsymbol{\zeta}\}$, the power allocation policy, power splitting policy, and subcarrier allocation policy for receiver $k$ using power splitting mode $n$ on subcarrier $i$ are given by
 \begin{eqnarray}\label{eqn:power2}
\hspace*{-5mm}\tilde{P}_{i,k}^{n*}\hspace*{-2mm}&=&\hspace*{-2mm}s_{i,k}^n{P}_{i,k}^{n*}=s_{i,k}
\Bigg[\frac{W(\alpha_k+\gamma+\upsilon_k)}
{\ln(2)\Big(\Phi_{i,k}\Big)}\Bigg]^+\hspace*{-3mm}, \,\forall i,k,\,\\
\label{eqn:a_selection}
   \hspace*{-5mm}{a}_{k}^{n*}\hspace*{-3mm}&=&\hspace*{-3mm}\left\{ \begin{array}{rl}
 1 &\mbox{if $n=\arg \underset{b}{\max} \quad T_k^b$ },\forall k, \\
 0 &\mbox{otherwise}
       \end{array} \right.\end{eqnarray}\begin{eqnarray}
       \hspace*{-4mm} T_{k}^b\hspace*{-1mm}&=&\hspace*{-1mm} \rho_{k}^{E_b} \Big( q\eta_k\sum_{i=1}^{n_F}(\sigma_{z^a}^2+
\sigma_{I_{i,k}}^2)\hspace*{-1mm}-\hspace*{-1mm}\sum_{i=1}^{n_F}\sum_{m=1}^{N}\delta_{i,k}^m -\sum_{i=1}^{n_F}\kappa_{i,k}^b \Big)\notag\\
&&- \frac{w^b_k P_{\min_k}^{\rm req}}{\rho_{k}^{E_b}}-\varphi_k,\\
  \label{eqn:sub_selection2} s_{i,k}^{n*}\hspace*{-3mm}&=&\hspace*{-3mm}
 \left\{ \begin{array}{rl}
 1 &\mbox{if $n,k=\arg \underset{c,b}{\max} \quad  \ M_{i,b}^c$ }, \forall i, \\
 0 &\mbox{ otherwise}
       \end{array} \right.\label{eqn:marginal2}\\
       M_{i,k}^n&=&W\Big (\alpha_k+\gamma+\upsilon_k\Big)\Big[\log_2\Big(P_{i,k}^{n*} l_k
g_k\abs{H_{i,k}}^2\Big)\notag\\
&&+\log_2\Big(\frac{\rho^{I_n}_{k}}
{\rho^{I_n}_{k}(\sigma_{z^a}^2+\sigma_{I_{i,k}}^2)+\sigma_{z^s}^2}\Big)\notag\\
&&-\frac{1}{\ln(2)}-\frac{\sigma_{z^s}^2}{\ln(2)(\rho^{I_n}_{i,k}(\sigma_{z^a}^2+
\sigma_{I_{i,k}}^2)+\sigma_{z^s}^2)}\Big]\notag\\
&&-\delta_{i,k}^n\rho_k^{I_n}-\Big(\sum_{m=1}^N \kappa_{i,k}^m\Big)\rho_k^{I_n}-\zeta_i
\end{eqnarray}
and $\Phi_{i,k}$ is defined in (\ref{eqn:phi}).
 The power allocation solution in (\ref{eqn:power2}) has a similar multi-level water filling interpretation as in (\ref{eqn:power1}).
The  difference between (\ref{eqn:power1}) and (\ref{eqn:power2}) is that the power allocation in (\ref{eqn:power2}) is performed w.r.t. each power splitting mode.  On the other hand, it can be observed from (\ref{eqn:a_selection}) and (\ref{eqn:sub_selection2})   that the optimal values of $a_{k}^n$ and  $s_{i,k}^n$  are binary numbers, although  time sharing relaxation is used for facilitating the algorithm design.

\begin{table*}[t]\caption{System parameters}\label{tab:feedback} \centering{
\begin{tabular}{|l|l|}\hline
\mbox{Maximum  service distance $d_{\max}$} & $10 \mbox{ m}$  \\
                            \hline
\mbox{Reference distance $d_0$} & $1 \mbox{ m}$  \\
                                                        \hline
 \mbox{Users distribution} & \mbox{Uniformly distributed between $d_0$ and $d_{\mathrm{\max}}$}  \\

    \hline
 \mbox{Multipath fading distribution} & \mbox{Rician  fading with
Rician factor $6$ dB }  \\
 \hline
 \mbox{Channel path loss model} & \mbox{TGn path loss model  \cite{report:tgn} }  \\

    \hline
\mbox{Carrier center frequency} & $470$ MHz  \\
  \hline
\mbox{Number of subcarriers $n_F$} &  $128$  \\
    \hline
    \mbox{Total bandwidth ${\cal B}$ and subcarrier bandwidth $W$} &  $20$ MHz and $156$ kHz  \\
    \hline
    \mbox{Thermal noise power and antenna noise power  per subcarrier} &  \mbox{$-112$ dBm and  $-115$ dBm}   \\
    \hline
      \mbox{Quantization noise (12-bit uniform quantizer)  per subcarrier} &  \mbox{$-47$ dBm}   \\
    \hline
        \mbox{Minimum system data rate requirement $R_{\min}$} &  $50$ \mbox{Mbit/s}  \\
    \hline
            \mbox{Minimum  data rate requirement for delay constrained service $R_{\min_k}$} &  $10$ \mbox{Mbit/s}  \\
    \hline
            \mbox{Class A/B power amplifier \cite{CN:Amplifier-wideband} with a power efficiency of
$16\%$ } & $\varepsilon=6.25$  \\
             \hline
            \mbox{Maximum power supply $P_{PG}$} & $50$ dBm \\
    \hline
            \mbox{Minimum required power transfer $P^{\rm req}_{\min_k}$} & $0$ dBm  \\
    \hline
                \mbox{Energy harvesting efficiency $\eta_k$} & $0.8$  \\
    \hline
            \mbox{Circuit power consumptions: $P_{C_T}$  and $P_{C_R}$}& $30$ dBm and $20$ dBm  \\
                \hline
                            $\mbox{Effective antenna gain}$ & $12$ dB  \\
                              \hline

\end{tabular}}
\end{table*}

Now, since the dual function is differentiable, we can update the set of Lagrange
multipliers
 for a given set of ${\cal P}^N$, ${\cal S}^N$, ${\cal A}^N$ by using the gradient
method. The gradient
update equations are given by
\begin{eqnarray}\label{eqn:multipler11}
\hspace*{-5.5mm}\lambda(u+1)\hspace*{-3mm}&=&\hspace*{-3mm}\Big[\lambda(u)-\xi_1(u)\times
\Big(  P_{\max}\notag\\
&-&\sum_{i=1}^{n_F}\sum_{k=1}^{K}\sum_{n=1}^{N}\tilde{P}_{i,k}^n\Big)\Big]^+\hspace*{-1.5mm},\\
\hspace*{-5.5mm}\beta(u+1)\hspace*{-3mm}&=&\hspace*{-3mm}\Big[\beta(u)-\xi_2(u)\times
\Big( P_{PG} -P_C\notag\\
&-&\sum_{i=1}^{n_F}\sum_{k=1}^{K}\sum_{n=1}^{N}\varepsilon \tilde{P}_{i,k}^n)\Big]^+\hspace*{-1.5mm}, \label{eqn:multipler22}\\
\hspace*{-5.5mm}\gamma(u+1)\hspace*{-3mm}&=&\hspace*{-3mm}\Big[\gamma(u)-\xi_3(u)\times
\Big(R_{\min}\notag\\
&-&\sum_{i=1}^{n_F}\sum_{k=1}^{K}\sum_{n=1}^{N}  s_{i,k}^n \widehat{C}_{i,k}^n\Big )\Big]^+\hspace*{-1.5mm},\label{eqn:multipler33}\\
\hspace*{-5.5mm}\delta_{i,k}^n(u+1)\hspace*{-3mm}&=&\hspace*{-3mm}\Big[\delta_{i,k}^n(u)-\xi_4(u)
\notag\\
&\times&\Big(1-\rho^{I_n}_{k} s_{i,k}^n- \sum_{m=1}^N\rho^{E_m}_{k}a_{k}^m\Big)\Big]^+,\forall i, k, n, \label{eqn:multipler44}\\\label{eqn:multipler55}
\hspace*{-5.5mm}\kappa_{i,k}^n(u+1)\hspace*{-3mm}&=&\hspace*{-3mm}\Big[\kappa_{i,k}^n(u)-\xi_5(u)\notag\\
&\times&
\Big(1-\sum_{m=1}^{N}\rho^{I_m}_{k} s_{i,k}^m- \rho^{E_k}_{k}a_{k}^n\Big)\Big]^+,\forall i, k,n ,\\
\hspace*{-5.5mm}w^n_{k}(u+1)\hspace*{-3mm}&=&\hspace*{-3mm}\Big[w^n_{k}(u)-\xi_6(u)\Big.\end{eqnarray}\begin{eqnarray}
\hspace*{-3mm}&&\times\Big.
\Big(\sum_{i=1}^{n_F} \Big(\sum_{j=1}^{K}\tilde{P}_{i,j}\Big) l_k g_k \abs{H_{i,k}}^2 \eta_k\notag\\
 \hspace*{-3mm}&&+ \sum_{i=1}^{n_F}(\sigma_{z^a}^2+
\sigma_{I_{i,k}}^2)  \eta_k - \frac{a^k_n P_{\min_k}^{\rm req}}{\rho_{k}^{E_n}}, \Big)\Big]^+,\forall  k,n, \notag \label{eqn:multipler77} \end{eqnarray} \begin{eqnarray}
\upsilon_k(u+1)\hspace*{-3mm}&=&\hspace*{-3mm}\Big[\upsilon_k(u)-\xi_7(u)\times
\Big(R_{\min_k}\notag\\
 \hspace*{-3mm}&&-\sum_{i=1}^{n_F}\sum_{n=1}^{N} s_{i,k}^n\widehat{C}_{i,k}^n\Big )\Big]^+, \forall k\in{\cal D}.\label{eqn:multipler88}
\end{eqnarray}
Similar to the case of continuous  power splitting ratios, updating $\zeta_i$ and $\varphi_i$  is
not necessary since they will not affect the power splitting mode selection  and subcarrier allocation
in (\ref{eqn:a_selection}) and (\ref{eqn:sub_selection2}), respectively.

\section{Results}
\label{sect:result-discussion}
In this section, simulation results are presented to demonstrate
the energy efficiency and system capacity of the proposed resource allocation algorithms. We consider an indoor communication system with $K$ receivers and the corresponding simulation parameters are provided in Table \ref{tab:feedback}. We note that the $470$ MHz frequency band will be used by IEEE 802.11 for the next generation of Wi-Fi systems  \cite{report:80211af}. Besides, the wavelength of the carrier signal is $0.6$ meter which is smaller than the minimum distance between the transmitter and receivers. Thus, the far-field assumption of the channel model in \cite{report:tgn} holds. Furthermore, in practice, the value of circuit power
consumption depends on the specific hardware implementation and the  application. On the other hand, we assume that all receivers have the same priority $\alpha_k=1,\,\forall k$,  for illustrating the maximum achievable energy efficiency of the system. The shadowing of all communication links is set to $g_k=1,\,\forall k$, to account for the line-of-sight communication setting.
 Unless specified otherwise, we assume that  there is only one receiver requiring delay constrained service with a minimum data rate requirement of $R_{\min_k}=10$ Mbit/s.  In the  sequel, the total number of iterations is defined as the number of main loop iterations  in the Dinkelbach method. For the case of continuous power splitting ratios, we set   $\rho_{U}^{E}=\rho_{U}^{I}=1 $ and  $\rho_{L}^{E}=\rho_{L}^{I}=0$. Besides, there are five power splitting ratios for the resource allocation with the discrete set of power splitting ratios: $\rho^{E}_{i,k}\in\{1,0.75,0.5,0.25,0\}$ and
$\rho^{I}_{i,k}\in\{0,0.25,0.5,0.75,1\}$.  Moreover, to ensure fast convergence, the step sizes adopted in (\ref{eqn:multipler1})--(\ref{eqn:multipler8}) and (\ref{eqn:multipler11})--(\ref{eqn:multipler88}) are optimized via backtracking line search,  cf.  \cite[page 464]{book:convex} and \cite[page 230]{book:non_linear_programming}.  Note that if the transmitter is
unable to meet the minimum required system data rate $R_{\min}$, the minimum required individual data rate  $R_{\min_k}$,  or
the minimum required power transfer $P^{\rm req}_{\min_k}$, we set the energy
efficiency and the system capacity for that channel realization
to zero to account for the corresponding failure. For the sake of illustration, the performance curves of the proposed algorithms for the continuous and discrete sets of  power splitting ratios are labeled as ``Proposed algorithm I" and ``Proposed algorithm II" in Figures \ref{fig:convergence}--\ref{fig:EE_K_CAP_K}. The average  energy efficiency of the system is computed according to (\ref{eqn:avg-sys-eff}) and averaged over 100000  independent  realizations of multipath fading and path loss attenuation.

\subsection{Convergence and Optimality of Iterative Algorithm }
\begin{figure}[t]
 \centering
\includegraphics[width=3.5 in]{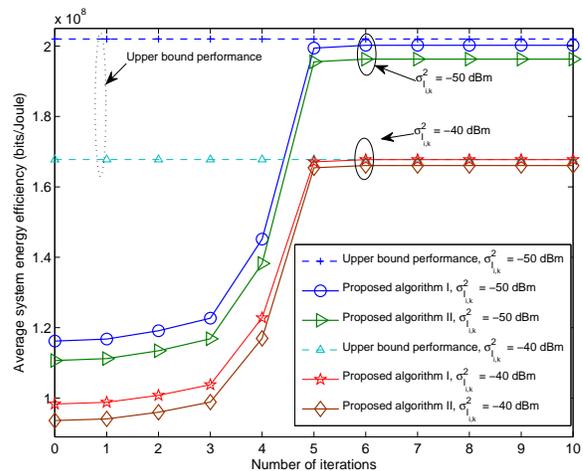}
\caption{Average system energy efficiency (bits-per-Joule) versus number of
iterations with different
 levels of interference power, $\sigma_{I_{i,k}}^2$, and  $P_{\max}=30$ dBm. There are $K=3$ receivers in the system.  The dashed
solid lines
represent the upper bound of  energy efficiency after algorithm convergence. } \label{fig:convergence}
\end{figure}

Figure \ref{fig:convergence} depicts the average system energy efficiency of  the
proposed iterative algorithms for different levels of  received interference versus the number of iterations. Specifically, we are interested in the energy efficiency  and  convergence speed of the proposed algorithms. We  plot the upper bound performance\footnote{The upper bound system performance is obtained by directly evaluating the upper bound objective function via replacing the energy consumption function $U_{TP}({\cal P},{\cal S},\boldsymbol\rho)$ in (\ref{eqn:U_TP-bound}) by  $\widehat{U}({\cal P},{\cal S},\boldsymbol\rho )$ for the continuous set of power splitting ratios. We note that the approximation in (\ref{eqn:cap-approx}) is asymptotically tight for high SINR.} of the system after convergence to illustrate the sub-optimality of the proposed algorithms.   The dashed
lines refer to the average  energy efficiency upper bound for each case study. After only 5 iterations,
 the iterative algorithms  achieve over 95\% of the
upper bound value for all considered scenarios. Besides, the convergence speed of the proposed algorithms is invariant to  the interference  levels, $\sigma_{I_{i,k}}^2$, which is desirable for practical implementation.

In the following case studies, the number of iterations is set to 5 for illustrating the performance of the proposed algorithms.

\subsection{Energy Efficiency versus Maximum Allowed Transmit Power}
\begin{figure*}[t]
 \centering
\includegraphics[angle = -90,width=6.5in]{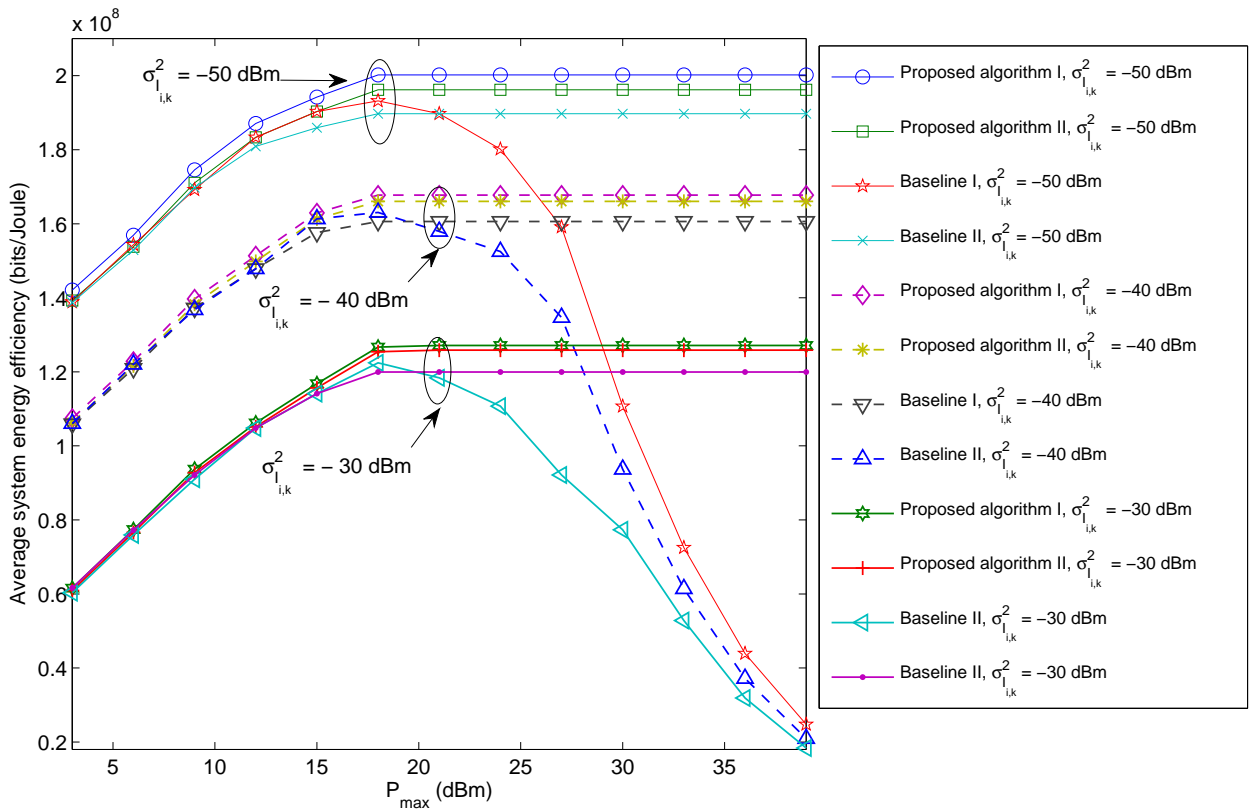}
\caption{Average system energy efficiency (bits-per-Joule) versus
maximum transmit power allowance, $P_{\max}$, for $K=3$ receivers and different levels of interference power, $\sigma_{I_{i,k}}^2$.} \label{fig:EE_PT}
\end{figure*}

Figure \ref{fig:EE_PT} shows the  average system energy efficiency versus the
maximum transmit power allowance, $P_{\max}$, for different received levels of interference, $\sigma_{I_{i,k}}^2$.  We first focus on the case of a small number of receivers and moderate interference level.  It can be observed that the average system energy efficiency of the proposed algorithms is a monotonically non-decreasing function of $P_{\max}$. In particular, starting from a small value of $P_{\max}$,   the energy efficiency first quickly increases with an increasing $P_{\max}$ and then saturates when $P_{\max}>18$ dBm. This is due to the fact that the two proposed algorithms strike a balance between the system energy efficiency and the power consumption of the system. In fact, once the maximum energy efficiency\footnote{The maximum energy efficiency refers to the ``maximum" w.r.t. the  corresponding problem formulation.} of the system is achieved, a further increase in the transmit power would result in a degradation in energy efficiency. As expected, proposed algorithm I outperforms proposed algorithm II in all cases since the latter algorithm is designed based on discrete sets of power splitting ratios which span a smaller feasible solution set compared to algorithm I. Besides, Figure \ref{fig:EE_PT} reveals that although interference signals can act  as a viable energy source  to the system, cf. (\ref{eqn:Power_harvested_d}) and (\ref{eqn:avg-sys-eff}), strong interference impairs the energy efficiency of the system; the energy harvesting gain  due to strong interference\footnote{Nevertheless, we would like to emphasize that the use of hybrid information and energy harvesting receivers provides a better energy efficiency to the system compared to  pure information receivers, as can be observed from  baseline II.} is unable to compensate the corresponding capacity loss.
  For comparison, Figure
\ref{fig:EE_PT} also contains the energy efficiency of two baseline
resource allocation schemes.  For baseline I, we maximize the weighted  system capacity
(bit/s) with respect to $\{{\cal P, S}, \boldsymbol\rho\}$ subject to constraints C1--C12 in (\ref{eqn:cross-layer}),
instead of the energy efficiency. Specifically, baseline I focuses on a system with hybrid energy harvesting receivers which split the received signal into two power streams for a finite discrete set of power splitting ratios. On the other hand,
  baseline II maximizes the weighted energy efficiency of the system with respect to $\{\cal P,S\}$ for receivers which do not have energy harvesting capability. Since the receivers of baseline II are unable to harvest energy, we do not impose constraint C1 for baseline II since  a minimum power transfer to the receivers is not required. Figure \ref{fig:EE_PT} reveals that in the low transmit power regime and for a small number of receivers, the system with  power splitting receivers achieves a small performance gain compared to the system without energy harvesting receivers. This is because in the low transmit power regime, the received power of the desired signal at the receivers may not be sufficiently large for simultaneous information decoding and energy harvesting. As a result, power splitting occurs for fulfilling constraint C1. On the other hand, the energy efficiency gain achieved by the proposed algorithm attains its maximum in the high transmit power allowance regime.  It can be observed that approximately a  5\% gain in energy efficiency can be achieved for $K=3$ receivers  and a moderate interference power level.  We  emphasize that in the high transmit power allowance regime, the maximum energy efficiency achieved by the system with hybrid energy harvesting receivers cannot be attained by the system without hybrid energy harvesting receivers (baseline II) via increasing the transmit power.

Now, we focus on the system with a moderate number of receivers and moderate interference levels in the high transmit power allowance regime.
\begin{figure*}[t]
\centering
\includegraphics[width=6.5in]{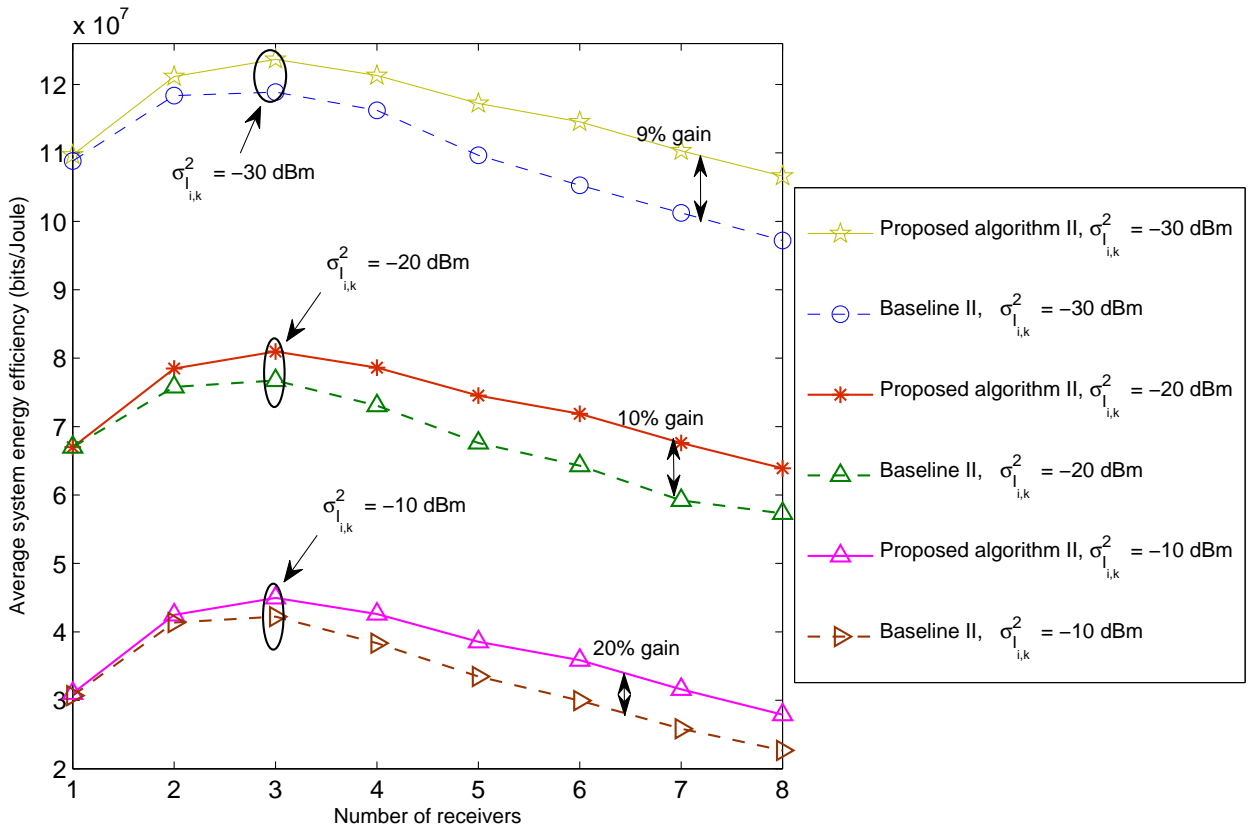}
\caption{Average system energy efficiency (bits-per-Joule) versus
number of receivers, $K$, for $P_{\max}=25$ dBm and different levels of interference power, $\sigma_{I_{i,k}}^2$.  }\label{figure4}
\end{figure*}
Figure \ref{figure4} illustrates the average system energy efficiency versus the number of receivers for different interference power levels, $\sigma_{I_{i,k}}^2$, and different resource allocation algorithms.  The maximum transmit power allowance is set to $P_{\max}=25$ dBm. In particular, we compare the proposed algorithm II, i.e., energy efficiency maximization with discrete power splitting ratio receivers, with baseline II. It can be observed from Figure \ref{figure4} that  the performance gain provided by energy harvesting depends on the number of receivers and the interference power level for a fixed maximum transmit power allowance $P_{\max}$. Indeed, when there is only one receiver, the performance gain due to energy harvesting is small since only a small portion of  radiated power can be harvested by the receiver. However, when there are more receivers in the systems, more receivers  participate in the energy harvesting process and thus a larger amount of energy can be harvested from the RF signals. As a result, a larger performance gain in terms of energy efficiency can be achieved by the proposed algorithm over baseline II when there are more receivers in the system. Furthermore, it can be seen that the performance gain achieved by the proposed algorithm over baseline II increases with the interference  power level.
 The reason behind this is twofold. First, the increase in interference power level provides some extra energy to the system for potential energy harvesting. Second, the strong interference tends to saturate the  SINR on each subcarrier such that it is independent of $\rho_{i,k}^{I*}$, i.e., $\frac{P_{i,k}\rho_{i,k}^{I*} l_k
g_k\abs{H_{i,k}}^2}{\rho_{i,k}^{I*}(\sigma_{z^a}^2+\sigma_{I_{i,k}}^2)+\sigma_{z^s}^2}\Rightarrow \frac{P_{i,k} l_k
g_k\abs{H_{i,k}}^2}{\sigma_{z^a}^2+\sigma_{I_{i,k}}^2} $ for $\sigma_{z^a}^2+\sigma_{I_{i,k}}^2\gg \sigma_{z^s}^2$. Thus, using more of the received power for information decoding does not provide a significant gain in channel capacity. Consequently, more received power is used for energy harvesting to reduce the total energy consumption of the system (see also Figure \ref{fig:power_harvested}), which enhances the system energy efficiency. On the other hand, we would like to emphasize that although the energy efficiency gain of algorithm II compared to baseline II depends on the system settings, energy harvesting may be necessary in some applications where other power sources are not available for the receivers regardless of how much can be gained in energy efficiency.
\begin{figure*}[t]
\centering
\includegraphics[width=6.5 in]{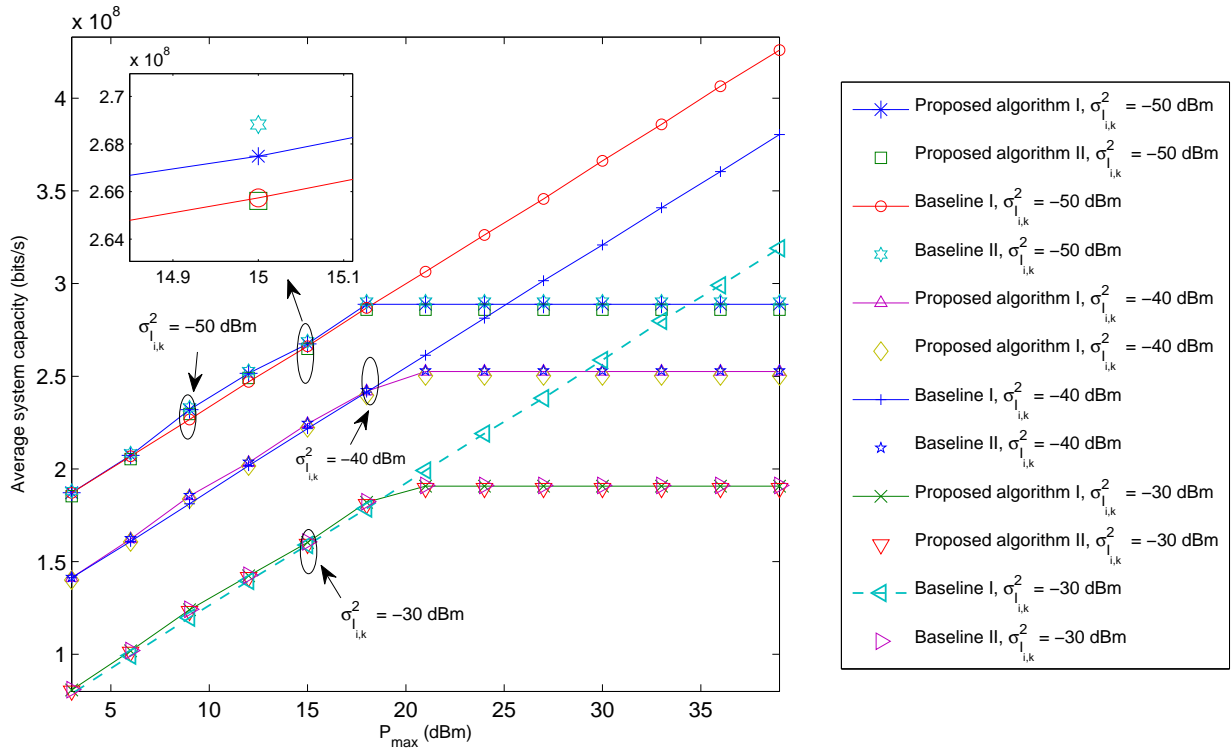}\vspace*{-10mm}
 \caption{Average system capacity (bits-per-second) versus
maximum transmit power allowance, $P_{\max}$, for different power  levels of interference,  $\sigma_{I_{i,k}}^2$, and $K=3$ receivers.} \label{fig:CAP_PT}
\end{figure*}

\subsection{Average System Capacity versus Maximum Allowed Transmit Power}

In Figure \ref{fig:CAP_PT}, we plot the average system capacity versus the
maximum transmit power allowance, $P_{\max}$, for different levels of interference power,  $\sigma_{I_{i,k}}^2$.  We compare the two proposed algorithms with the two aforementioned baseline schemes. For $P_{\max}<18$ dBm,  it
can be observed that the average system capacities of the two proposed
algorithms scale with the maximum transmit power allowance $P_{\max}$. Yet, the system capacity gain due to a larger $P_{\max}$ begins to saturate in the high transmit power allowance regime, i.e.,  $P_{\max}\ge 18$ dBm. Indeed, the proposed algorithms do not further increase the transmit power in the RF if the system capacity gain due to a higher transmit power cannot  neutralize the associated energy consumption increase required for boosting the transmit power.  Baseline II has a similar behaviour as the proposed algorithms with respect to the maximum transmit power allowance, since baseline II also focuses on energy efficiency maximization.  On the other hand, as expected, baseline  II achieves a system capacity gain over the other algorithms in the low transmit power allowance regime. Indeed, the SINR of each subcarrier is monotonically increasing with respect to $\rho_{i,k}^I$. In other words, the SINR (capacity) on each subcarrier is maximized when there is no power splitting, i.e., $\rho_{i,k}^I=1,\rho_{i,k}^E=0$, which is the setting for baseline II. We note that although baseline I focuses on system capacity maximization, power splitting occurs for fulfilling constraint C1 which results in a lower system capacity compared to baseline II in the low transmit power regime. However, the system capacity of  baseline I is higher than that of the other algorithms in the high transmit power regime, e.g., $P_{\max}\ge 18$ dBm. This is attributed to the fact that in order to maximize the system capacity for baseline I, the transmitters always radiate all the available power. Yet, the higher
system capacity of baseline I comes at the expense of a low system energy efficiency, cf. Figure \ref{fig:EE_PT}.
 In addition, we observe from Figure \ref{fig:CAP_PT} that all the considered algorithms are able to fulfill the system data rate requirement in C4 on average.

\begin{figure}
 \centering
\includegraphics[width=3.5in]{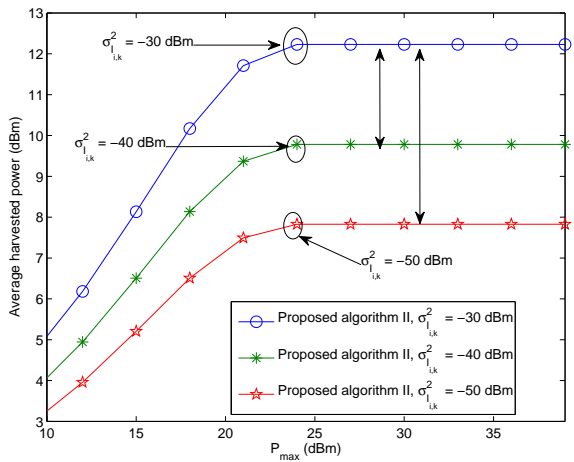}
\caption{Average harvested power (dBm) of proposed algorithm II versus
the maximum transmit power allowance, $P_{\max}$,  for different interference power levels and $K=3$ receivers. The double-sided arrow
indicates the power harvesting gain due to an increasing interference power level. } \label{fig:power_harvested}
\end{figure}

\subsection{Average Harvested Power versus Maximum Allowed Transmit Power}
Figure \ref{fig:power_harvested}  depicts the average harvested power of the proposed algorithm II versus the
maximum allowed transmit power, $P_{\max}$, for different levels of interference power,  $\sigma_{I_{i,k}}^2$.
It can be seen that in the high transmit power regime, the amount of average harvested power in all considered scenarios is saturated. This is because for the proposed algorithm II, the transmitter stops to increase the transmit power for energy efficiency maximization. Meanwhile, the average interference power level remains unchanged  and no extra energy can be harvested by the $K$ receivers. On the other hand, a higher amount of power is harvested by the receivers in the system when the interference power level increases as the SINR is independent of $\rho_{I_{i,k}}^k$ on each subcarrier. Thus, splitting more received power for energy harvesting improves system energy efficiency and increases the average harvested power.

\begin{figure*}[t]
\subfigure[Average energy efficiency (bits-per-Joule) versus  $K$.]{
\includegraphics[width=3.5 in]{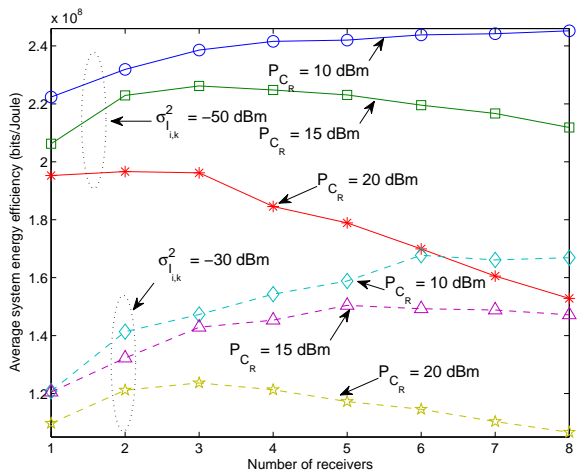}
\label{fig:ee_k} }\centering \subfigure[ Average system capacity (bit-per-second) versus  $K$.]{
\includegraphics[width=3.5 in]{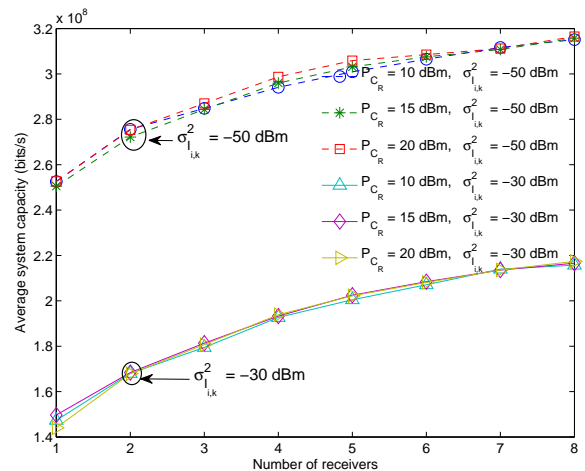}
\label{fig:cap_k} }\caption[Optional caption for list of figures]
{Average energy efficiency (bits-per-Joule) and average system capacity (bit-per-second) versus the number of receivers $K$ for $P_{\max}=18$ dBm,
  different receiver circuit power consumptions, $P_{C_R}$, and different interference  power levels $\sigma_{I_{i,k}}^2$.}\label{fig:EE_K_CAP_K}
\end{figure*}

\subsection{Average Energy Efficiency and System Capacity versus Number of Receivers}
 Figure   \ref{fig:ee_k} and  Figure \ref{fig:cap_k} illustrate the average system capacity and the  average system energy efficiency  of the proposed algorithm II versus the number of mobile receivers for different interference power levels, $\sigma_{I_{i,k}}^2$, and different receiver circuit power consumptions, $P_{C_R}$.   In  Figure \ref{fig:cap_k}, it can be observed that the average system capacity increases with the number of receivers in the system since the proposed algorithm is able to exploit multiuser diversity. Specifically, the transmitter has a higher chance of selecting a receiver with good channel conditions when more receivers are in the system, which results in a system capacity gain. In addition, although a higher interference power level impairs the average system capacity,  it does not decrease the performance gain due to multiuser diversity as can be concluded from the slopes of the curves. Besides,  a higher  circuit power consumption in the receivers does not have a large impact on the average system capacity.   On the contrary, Figure  \ref{fig:ee_k} shows that the average system energy efficiency does not necessarily monotonically increase/decrease  w.r.t. the number of receivers. In fact, for a moderate value  of  receiver circuit power consumption, e.g. $P_{C_R}\ge 15$  dBm, for the considered system setting, the energy efficiency of the system   first increases and then decreases with an increasing  number of receivers. The enhancement of energy efficiency  is  mainly due to the multiuser diversity gain in the channel capacity when having multiple receivers.  Besides, if more receivers  participate
in the energy harvesting process, a larger portion of energy can be harvested from the RF signals. Nevertheless, an extra circuit energy consumption  is incurred by each additional receiver. Indeed, when the number of receivers in the system or $P_{C_R}$  are large, the system performance gain due to multiuser diversity is unable to compensate the  total energy cost of  the receivers since $K P_{C_R}$ increases linearly w.r.t. the number of receivers. Hence, the energy efficiency of the system decreases with the number of receivers.
 As a matter of fact,  the energy efficiency gain due to the additional receivers depends on the trade-off between  multiuser diversity gain, the amount of harvested energy,  and the associated cost in having multiple receivers in the system. In the extreme case, when $P_{C_R}\rightarrow 0$, the energy efficiency will monotonically increase w.r.t. the number of receivers, provided that the optimization problems in (\ref{eqn:cross-layer}) and (\ref{eqn:cross-layer-discrete}) are feasible.

\section{Conclusions}\label{sect6}
In this paper, the resource allocation
algorithm design for simultaneous  wireless information and power transfer in OFDMA systems was studied.
We focused on power splitting receivers which are able to split the received signals into two power streams for concurrent information decoding and energy harvesting. The algorithm design was formulated as a non-convex optimization problem which  took into account a minimum system data rate
requirement, minimum individual data rate
requirements of the receivers, a minimum required power transfer, and  the total system power
dissipation. We first focused on receivers with continuous sets  of power splitting ratios and proposed a resource allocation algorithm. The
derived  solution served as a building block for the design of a suboptimal resource allocation algorithm for receivers with discrete sets of power splitting ratios. Simulation results
showed the excellent performance of the two proposed suboptimal algorithms
and also unveiled the trade-off between energy efficiency, system capacity, and wireless power transfer. In particular, wireless power transfer enhances the system energy efficiency by harvesting energy in the radio frequency, especially in the interference limited regime.


\section*{appendix}

\subsection{Proof of Quasi-Concavity of the  Objective Function with respect to the Power Allocation Variables} \label{appendix:concave}

For facilitating the following presentation, we first provide a definition for quasi-concave functions. Consider a function $g(\cdot)$
with input vector $\mathbf{x}$.  $g(\mathbf{x})$ is a strictly quasi-concave function if and only if its
super-level set \cite{book:convex}
\begin{eqnarray}
L_\nu= \{\mathbf{x}\in\mathbf{R}^n | g(\mathbf{x})\ge \nu\},\,\,\, \nu\in \mathbf{R},
\end{eqnarray}
is concave, where  $\mathbf{R}$ denotes the set of real numbers.  Suppose $g({\cal P},{\cal S},\boldsymbol\rho)=\frac{ {U}({\cal P},{\cal S},\boldsymbol\rho )}{  {U}_{TP}({\cal P},{\cal S},\boldsymbol\rho )}$. Now, we focus on the following set:
\begin{eqnarray}
\frac{ {U}({\cal P},{\cal S},\boldsymbol\rho )}{  {U}_{TP}({\cal P},{\cal S},\boldsymbol\rho )}\ge  \nu. \label{eqn:superlevel-set}
\end{eqnarray}
When $\nu <0$, the set is empty since $ {U}({\cal P},{\cal S},\boldsymbol\rho )> 0$ and $ {U}_{TP}({\cal P},{\cal S},\boldsymbol\rho )>0$.  As a result, $L_\alpha$ is an affine(/concave) set  \cite{book:convex}  for $ \nu<0$.  Then, for $ \nu>0$, we can rewrite set $L_\nu$ as
\begin{eqnarray}\label{eqn:superlevel-set2}\notag
&&\frac{ {U}({\cal P},{\cal S},\boldsymbol\rho )}{  {U}_{TP}({\cal P},{\cal S},\boldsymbol\rho )}\ge \nu\\
\Longleftrightarrow && {U}({\cal P},{\cal S},\boldsymbol\rho )-\nu {U}_{TP}({\cal P},{\cal S},\boldsymbol\rho )\ge 0.
\end{eqnarray}

Since $ {U}({\cal P},{\cal S},\boldsymbol\rho )>0$ and $ {U}_{TP}({\cal P},{\cal S},\boldsymbol\rho )>0$ are concave and affine functions with respect to the power allocation variables ${\cal P}$, respectively,  (\ref{eqn:superlevel-set2}) is a concave function and $L_\nu$ represents a concave set for $ \nu>0$.
Therefore, the objective function is a strictly quasi-concave function with respect to the power allocation variables. In other words, the objective function is either 1) first monotonically  non-decreasing  and  then monotonically non-increasing or 2) monotonically  non-decreasing   with respect to the power allocation variables.

In the following, we show that the objective function is  first monotonically  non-decreasing  and  then monotonically non-increasing with respect to the power allocation variables. For facilitating the following analysis, we assume that
subcarrier $i$ is allocated to receiver $k$, i.e., $s_{i,k}^*=1$, while  $\rho_{i,k}^{E*}$ and $\rho_{i,k}^{I*}$  are given. Then, we take a partial derivative of the objective function with respect to ${P}_{i,k}$ which yields
\begin{eqnarray}\frac{\partial g({\cal P},{\cal S},\boldsymbol\rho)}{\partial {P}_{i,k}}&=& \frac{A_{i,k}}{B_{i,k}}\\
\mbox{where }A_{i,k}\hspace*{-1mm}&=&\hspace*{-1mm}{\Gamma}_{i,k} \Big(\Omega_{i,k} {P}_{i,k}+P_{C_T} + K P_{C_R}-\sum_{k=1}^K Q_{I,k}\Big)\notag\\
&&-\Omega_{i,k}(1+ {P}_{i,k}
   {\Gamma}_{i,k}) \ln (1+{P}_{i,k} {\Gamma}_{i,k})\notag,\\
   B_{i,k}\hspace*{-1mm}&=&\hspace*{-1mm}\ln(2) ({P}_{i,k} {\Gamma}_{i,k}+1)\notag\\
   &\times&\hspace*{-3mm}
   \Big(\Omega_{i,k} {P}_{i,k}+P_{C_T} + K P_{C_R}-\sum_{k=1}^K Q_{I,k}\Big)^2\hspace*{-1mm}, \mbox{and}\notag\\
  \Omega_{i,k}&=& \varepsilon-\sum_{k=1}^K l_k g_k \abs{H_{i,k}}^2\rho_{i,k}^{E*}.
\end{eqnarray}
We note that $\Omega_{i,k}>0$  holds since ${P}_{i,k}\ge\varepsilon{P}_{i,k}>{P}_{i,k}\sum_{k=1}^K l_k g_k \abs{H_{i,k}}^2 $ where the strict inequality is due to the second law of thermodynamics  from physics.
Then, we have
\begin{eqnarray}\label{eqn:p-small}
\frac{\partial g({\cal P},{\cal S},\boldsymbol\rho)}{\partial {P}_{i,k}}\Big|_{{P}_{i,k}\rightarrow 0}=-\frac{W{\Gamma}_{i,k}}{\sum_{k=1}^K Q_{I_k}-P_{C_T} -K P_{C_R}}>0
\end{eqnarray}
 due to $P_{C_T} + K P_{C_R}>\sum_{k=1}^K Q_{I_k}$.  In other words, the objective function is increasing when  ${P}_{i,k}$ increases from zero. On the other hand, when ${P}_{i,k}$ is sufficiently large, e.g., ${P}_{i,k}\rightarrow \infty$, the first partial derivative of the objective function with respect to ${P}_{i,k}$ is given by
\begin{eqnarray}\label{eqn:p-large}&&\hspace{-0.5cm}\notag\frac{\partial g({\cal P},{\cal S},\boldsymbol\rho)}{\partial {P}_{i,k}}\Big|_{{P}_{i,k}\rightarrow \infty}= \lim_{{P}_{i,k}\rightarrow \infty}\frac{A_{i,k}}{B_{i,k}}
\notag\\
&&\hspace*{-0.7cm}\stackrel{(a)}{=}\hspace*{-1.5mm}\lim_{{P}_{i,k}\rightarrow \infty}\hspace*{-0.5mm}\frac{-\Omega_{i,k}{\Gamma}_{i,k}\log_2(1+{P}_{i,k}{\Gamma}_{i,k})}{ ({\Gamma}_{i,k} (3 \Omega_{i,k}
  {P}_{i,k}\hspace*{-0.5mm}+\hspace*{-0.5mm}P_{C_T}\hspace*{-0.8mm} + \hspace*{-0.5mm}K P_{C_R}\hspace*{-1mm}-\hspace*{-1mm}\sum_{k=1}^K Q_{I_k}\hspace*{-0.8mm})\hspace*{-0.5mm}+\hspace*{-0.8mm}2 \Omega_{i,k})}\notag\\
  &&\times\frac{1}{(\Omega_{i,k} {P}_{i,k}+P_{C_T} + K P_{C_R}\hspace*{-1mm}-\hspace*{-0.5mm}\sum_{k=1}^K Q_{I_k}) }\notag\\
 &&\hspace*{-0.7cm} \le 0,
\end{eqnarray}
where $(a)$ is due to the use of  L'Hospital's rule.  Thus, the objective function is monotonic non-increasing for large increasing\ ${P}_{i,k}$. By combining (\ref{eqn:p-small}) and (\ref{eqn:p-large}), we conclude that the objective function is a quasi-concave function with respect to the power allocation variables. Specifically, it is first monotonically  non-decreasing  and  then monotonically non-increasing with respect to ${P}_{i,k}$.

\subsection{Proof of Concavity of the Transformed Problem  with Objective Function Approximation} \label{appendix:concave}
The concavity of the optimization problem with approximated objective function can be proved by the following few steps.
First, we consider the concavity of function $\widehat {U}({\cal P},{\cal S},\mbox{\boldmath$\rho$} )$ on a per
subcarrier basis w.r.t. the optimization variables $\tilde{P}_{i,k}$,
$\tilde{\rho}^I_{i,k}$, and  $\rho^E_{i,k}$. For notational simplicity, we define a vector $\mathbf{x}_{i,k}=[\tilde{P}_{i,k}\,\,\, \tilde{\rho}^I_{i,k} \,\,\, \rho^E_{i,k}]$ and a function
$f_{i,k}(\mathbf{x}_{i,k})= W\alpha_k \log_2\Big(\frac{\tilde{P}_{i,k}\tilde{\rho}^I_{i,k} l_k
g_k\abs{H_{i,k}}^2}{\tilde{\rho}^I_{i,k}(\sigma_{z^a}^2+\sigma_{I_{i,k}}^2)+\sigma_{z^s}^2}\Big)$ which takes vector $\mathbf{x}_{i,k}$ as input. Then, we denote by
$\mathbf{H}(f_{i,k}(\mathbf{x}_{i,k}))$, and $\tau_1$, $\tau_2$, and $\tau_3$ the Hessian matrix of function $f_{i,k}(\mathbf{x}_{i,k})$
and the eigenvalues of $\mathbf{H}(f_{i,k}(\mathbf{x}_{i,k}))$, respectively. The
Hessian matrix of function $f_{i,k}(\mathbf{x}_{i,k})$ is given by
\begin{eqnarray}&&\mathbf{H}(f_{i,k}(\mathbf{x}_{i,k}))\notag\\
&=&
\begin{bmatrix}
  \frac{-1}{(\tilde{P}_{i,k})^2 \ln(2)}&   0 &   0  \\
    0&    \frac{-\sigma_{z^a}^2\, \left(\sigma_{z^a}^2 + 2\, \tilde{\rho}^I_{i,k}\, \sigma_{I_{i,k}}^2\right)}{(\tilde{\rho}^I_{i,k})^2\, \ln(2) {\left(\sigma_{z^a}^2 + \tilde{\rho}^I_{i,k}\, \sigma_{I_{i,k}}^2\right)}^2}  & 0\\
    0 & 0 & 0
\end{bmatrix}
\end{eqnarray} and the corresponding eigenvalues are
$
\tau_1=\frac{-1}{(\tilde{P}_{i,k})^2 \ln(2)},\, \tau_2= \frac{-\sigma_{z^a}^2\, \left(\sigma_{z^a}^2 + 2\, \tilde{\rho}^I_{i,k}\, \sigma_{I_{i,k}}^2\right)}{(\tilde{\rho}^I_{i,k})^2\, \ln(2) {\left(\sigma_{z^a}^2 + \tilde{\rho}^I_{i,k}\, \sigma_{I_{i,k}}^2\right)}^2}, \mbox{ and  } \tau_3 =0.
$
Since $\tau_a\le 0, a\in\{1,2,3\}$, $\mathbf{H}(f_{i,k}(\mathbf{x}_{i,k}))$ is a negative semi-definite matrix. In other words, function $f_{i,k}(\mathbf{x}_{i,k})$ is jointly concave w.r.t.  $\tilde{P}_{i,k}$,
$\tilde{\rho}^I_{i,k}$, and  $\rho^E_{i,k}$. Then, we can perform the perspective transformation on $f_{i,k}(\mathbf{x}_{i,k})$ which is given by
$u_{i,k}(\mathbf{x}_{i,k})=s_{i,k}f_{i,k}(\mathbf{x}_{i,k}/s_{i,k})$. We note that the perspective transformation preserves the concavity of the function \cite{book:convex} and $u_{i,k}(\mathbf{x}_{i,k})$ is jointly concave w.r.t.  $\tilde{P}_{i,k}$,
$\tilde{\rho}^I_{i,k}$,   $\rho^E_{i,k}$, and $s_{i,k}$. Subsequently, $\widehat {U}({\cal P},{\cal S},\mbox{\boldmath$\rho$} )=\sum_{i=1}^{n_F}\sum_{k=1}^K W\alpha_k\, u_{i,k}(\mathbf{x}_{i,k}) $ can be constructed as a non-negative weighted sum of $u_{i,k}(\mathbf{x}_{i,k})$ which guarantees the concavity of the resulting function. Besides, $ \widehat {U}_{TP}({\cal P},{\cal S},\mbox{\boldmath$\rho$})$ is an affine function of the optimization variables. Thus, $\widehat {U}({\cal P},{\cal S},\mbox{\boldmath$\rho$} )-q \widehat {U}_{TP}({\cal P},{\cal S},\mbox{\boldmath$\rho$})$ is jointly concave w.r.t. the optimization variables. On the other hand,  constraints C1--C11 (with relaxed constraint C7), C14, and C15 span a convex feasible set. As a result, the transformed problem with the approximated objective function is a concave maximization problem.

\bibliographystyle{IEEEtran}
\bibliography{OFDMA-AF}

\end{document}